\documentclass[structabstract]{aa}  
\usepackage{graphicx}
\usepackage{txfonts}
\def\Msun{M$_{\sun}$}
\def\Mjup{M$_{\rm Jup}$}
\newcommand{\excs}{\extracolsep{\fill}}
\begin{document}

\title{Deep imaging survey of young, nearby austral stars}

\subtitle{VLT/NACO Near-Infrared Lyot-Coronographic Observations}  

\author{
        G. Chauvin\inst{1}\and
        A.-M. Lagrange\inst{1}\and
	M. Bonavita\inst{2, 3}\and
        B. Zuckerman\inst{4}\and
        C. Dumas\inst{5}\and
	M.\,S. Bessell\inst{6}\and
        J.-L. Beuzit\inst{1}\and
        M. Bonnefoy\inst{1}\and
	S. Desidera\inst{2}\and
	J. Farihi\inst{7}\and
        P. Lowrance\inst{8}\and
        D. Mouillet\inst{1}\and
	I. Song\inst{9}
}   

\institute{
        Laboratoire d'Astrophysique, Observatoire de Grenoble, UJF, CNRS:
        414, Rue de la piscine, 38400 Saint-Martin d'H\`eres, France
        \and
	INAF - Osservatorio Astronomico di Padova, Vicolo dell' Osservatorio 5, 35122 Padova, Italy
        \and
	Universita' di Padova, Dipartimento di Astronomia, Vicolo dell'Osservatorio 2, 35122 Padova, Italy
	\and
	Department of Physics \& Astronomy and Center for Astrobiology, 
	University of California: Los Angeles, Box 951562, CA 90095, USA
        \and
	European Southern Observatory: Casilla 19001, Santiago 19, Chile
        \and
	Research School of Astronomy and Astrophysics Institute of Advance Studies, 
	Australian National University: Cotter Road, Weston Creek, Canberra, ACT 2611, Australia
        \and
	Department of Physics \& Astronomy, University of Leicester, Leicester LE1 7RH, United Kingdom
	\and
	Spitzer Science Center, IPAC/Caltech: MS 220-6, Pasadena, CA 91125, USA
        \and
	Department of Physics \& Astronomy, University of Georgia, Athens, GA 30602-2451, USA  
}

\date{Received September, 2008}

%
% \abstract{}{}{}{}{} 
% 5 {} token are mandatory
% 
  \abstract
   {High contrast and high angular resolution imaging is the optimal technique to
   search for substellar companions to nearby stars at physical
   separations larger than typically 10~AU. Two distinct populations of
   substellar companions, brown dwarfs and planets, can be probed and
   characterized. Fossile traces of their different formation
   processes should be revealed by their respective physical and
   orbital properties and should then allow testing basic aspects of
   their respective formation and evolution mechanisms.}
%   Detection
%   performances down to the planetary mass regime are now achievable
%   around members of young, nearby associations and two distinct
%   populations of substellar companions, brown dwarfs and planets, can
%   be probed and characterized. Fossile traces of their different
%   formation processes should be revealed by their respective physical
%   and orbital properties and should allow testing basic aspects of
%   their formation and evolution mechanisms.}
   {Since November 2002, we have conducted the largest deep imaging
   survey of the young, nearby associations of the southern
   hemisphere. Our goal is detection and characterization of
   substellar companions at intermediate ($10-500$~AU) physical
   separations.  We have observed a sample of 88 stars, mostly G to M
   dwarfs, that we essentially identify as younger than 100~Myr and
   closer to Earth than 100~pc. }
%   They are mainly members of the TW Hydrae, $\beta$
%   Pictoris, Tucana-Horologium and AB Doradus moving groups, but also
%   young stellar candidates currently not associated to any of these
%   groups.}
   {The VLT/NACO adaptive optics instrument of the ESO Paranal
   Observatory was used to explore the faint circumstellar environment
   between typically 0.1 and $10~\!''$. Diffraction-limited
   observations in $H$ and $K_s$-band combined with Lyot-coronagraphy
   enabled us to reach contrast performances as small as $10^{-6}$.
   The existence of planetary mass companions could therefore be
   probed. We used a standardized observing sequence to precisely measure
   the position and flux of all detected sources relative to their
   visual primary star. Repeated observations at several epochs
   enabled us to discriminate comoving companions from contaminants.}
   {We report the discovery of 17 new close ($0.1-5.0~\,''$) multiple
   systems. HIP\,108195\,AB and C (F1\,III-M6), HIP\,84642\,AB
   ($a\sim14$~AU, K0-M5) and TWA22\,AB ($a\sim1.8$~AU; M6-M6) are
   confirmed comoving systems. TWA22\,AB is likely to be a rare
   astrometric calibrator that can be used to test evolutionary model
   predictions.  Among our complete sample, a total of 65 targets were
   observed with deep coronagraphic imaging. About $240$ faint
   companion candidates were detected around 36 stars. Follow-up
   observations with VLT or HST for 83\% of these stars enabled us to
   identify a large fraction of contaminants. The latest results
   about the substellar companions to GSC\,08047-00232, AB\,Pic and
   2M1207, confirmed during this survey and published earlier, are
   reviewed. Finally, the statistical analysis of our complete set of
   coronagraphic detection limits enables us to place constraints on
   the physical and orbital properties of giant planets between
   typically 20 and 150~AU.}

   {}


   \maketitle

%
%________________________________________________________________

\section{Introduction}

The search for substellar objects, isolated, multiple or companion to
nearby stars, has been an important driver for observers in the two
last decades. Their detection and characterization contribute to
developing our understanding of the formation and evolution of stars,
brown dwarfs and planets. Since the discovery of the first unambiguous
brown dwarf Gl229\,B (Nakajima et al. 1995), the development of new instruments
and observing techniques has diversified. Large surveys (2MASS,
Skrutskie et al. 1997; DENIS, Epchtein et al. 1997; SLOAN, York et
al. 2000) are the best method for the study of isolated substellar
objects. Hundreds of brown dwarfs have been discovered in the field
motivating the introduction of the new L and T spectral classes
(Delfosse et al. 1997; Kirkpatrick et al 1999; Burgasser
et al. 1999).  Dedicated spectroscopic observations offer the
opportunity to study the physical and chemical processes of these very
cool atmospheres, such as grain and molecule formation, vertical
mixing and cloud coverage. In the field, in young open clusters or in
star forming regions, the study of the intial-mass function and of 
stellar and substellar multiplicity shows an apparent continuous
sequence supporting the idea that stellar mechanisms (collapse, fragmentation,
ejection, photo-evaporation of accretion envelopes) form objects over a wide range of masses,
down to planetary masses predicted by theoretical models
(Bonnell et al. 2007; Burgasser et al. 2007; Zuckerman \& Song
2009). Despite limited spatial resolution, a dozen substellar
companions to nearby stars have been discovered at wide ($\ge100$~AU)
orbits (Goldman et al. 1999, Kirkpatrick et al. 2000, Wilson
et al. 2001).

To access the near ($\le 5$~AU) environment of stars, other observing
techniques (radial velocity, transit, micro-lensing, pulsar-timing), 
are so far best suited. The radial velocity (RV)
and transit techniques are nowadays the most successful methods for
detecting and characterizing the properties of exo-planetary
systems. The RV surveys have focused on main sequence solar-type
stars, with numerous narrow optical lines and low activity, to ensure
high RV precision. Recently, planet-search programs have been extended
to lower and higher mass stars (Endl et al. 2006, Lagrange et
al. 2009) and younger and more evolved systems (Joergens et al. 2006,
Johnson et al. 2007). Since the discovery of 51 Peg\,b (Mayor \&
Queloz 1995), more than 300 exo-planets have been identified featuring
a broad range of physical (mass) and orbital (P, $e$) characteristics
(Udry \& Santos 2007; Butler at al. 2006). This technique also
revealed the existence of the so-called brown dwarf desert at small
($\le 5$~AU) separations (Grether \& Lineweaver 2006). The
bimodal aspect of the secondary mass distribution indicates different
formation mechanisms for two populations of substellar companions,
brown dwarfs and planets. The transit technique coupled with RV
enables determination of the radius and density of giant planets and
thus a probe of their internal structure. Moreover, spectral elements
of a planetary atmosphere can be revealed during primary or secondary
eclipse (Swain et al. 2008, Grillmair et al. 2008).

To extend such systematic characterization at larger scales
($\ge10$~AU), the deep imaging technique is particularly well suited
to probe the existence of planets and brown dwarf companions and
complete our view of planetary formation and evolution. To access
small angular separations, the space telescope (HST) or the
combination of Adaptive Optics (AO) system with very large
ground-based telescopes (Palomar, CFHT, Keck, Gemini, Subaru, VLT)
have become mandatory. Moreover, deep imaging surveys take advantage
of exhaustive work on identification of young ($\le100$~Myr), nearby
($\le 100$~pc) stellar associations.  Due to their youth and
proximity, such stars offer an ideal niche for detection of warm
planetary mass companions that are still moderately bright at
near-infrared wavelengths.  Since the recognition of the TW Hydrae
association (TWA; Kastner et al. 1997; Webb et al 1999), more than 200
young, nearby stars have been identified.  Many such stars reside in
several coeval moving groups (e.g., TWA, $\beta$ Pictoris, Tucana-Horologium,
$\eta$ Cha, AB Dor, Columba and Carinae), sharing common kinematics,
photometric and spectroscopic properties (see Zuckerman \& Song 2004, 
hereafter ZS04;
Torres et al. 2008, T08). A few young brown dwarf companions have been
detected from space, HR\,7329\,B and TWA5\,B (Lowrance et al. 2000, 1999), and from the ground, GSC\,08047-00232\,B
(Chauvin et al. 2005a). Companions down to the planetary mass regime were discovered
around the star AB Pic (Chauvin et al. 2005c) and the young brown dwarf 2M1207
(Chauvin et al. 2004, 2005b). Various deep imaging surveys of young, nearby
stars have recently been completed using different high contrast
imaging techniques such as coronagraphy, differential imaging or
$L$-band imaging (see Table~\ref{tab:deepsurveys}).  A significant
number have reported a null-detection result of substellar
companions. Kasper et al. (2007), Lafreni\`ere et al. (2007) and
Nielsen et al. (2008) have initiated a statistical analysis to
constrain the physical and orbital properties (mass, period,
eccentricity distributions) of a giant planet population.  Despite the
model-dependency on the mass predictions, the approach is attractive
for exploiting the complete set of detection performances of the survey and characterizing the outer
portions of exo-planetary systems.

\begin{table}[t]
\label{tab:deepsurveys}
\caption{Deep imaging surveys of young ($<100$~Myr), nearby
($<100$~pc) stars dedicated to the search for planetary mass
companions and published in the literature. The telescope and the
instrument (Tel/Instr.), the imaging mode (CI: coronagraphic imaging;
Sat-DI; saturated direct imaging; DI direct imaging; SDI: simultaneous
differential imaging; ADI: angular differential imaging) and filters,
the field of view (FoV) and the number of stars observed (\#) are
given. The typical survey sensitivity in terms of mass is also
reported with the survey reference.}
\begin{tabular*}{\columnwidth}{@{\excs}llllll}     % 7 columns 
\hline\noalign{\smallskip}
Tel/Instr.       &  Mode        & FoV          & \#   & Mass      &  Ref.     \\ 
                 & \& Filter    & (arcsec)     &      & (\Mjup)   &          \\
\noalign{\smallskip}\hline\noalign{\smallskip}
3.6m/ADONIS   & CI, $H-K$       & $13\times13$ & 29       & 5        & (1)                 \\ 
NTT/Sharp        & Sat-DI, $K$    & $11\times11$ & 23       & 5        & (2)                 \\ 
NTT/Sofi         & Sat-DI, $H$    & $13\times13$ & 10       & 5        & (2)                 \\ 
HST/NICMOS       & DI, H        & $19\times19$ & 45       & 1        & (3)                 \\
VLT/NaCo         & Sat-DI, $H-K$   & $14\times14$ & 28       & 5        & (4)                 \\ 
VLT/NaCo         & SDI, $H$       & $5\times5$   & 45       & 1        & (5)                 \\ 
VLT/NaCo         & DI, $L'$       & $28\times28$ & 22       & 1        & (6)                 \\ 
Gemini/NIRI      & ADI, $H$       & $22\times22$ & 85$^*$   & 1        & (7)                 \\
\noalign{\smallskip}\hline                  
\end{tabular*}
\begin{list}{}{}
\item[\scriptsize{- REFERENCES:}] \scriptsize{(1) Chauvin et al. 2003, (2) Neuh\"auser et al. 2003, (3) Lowrance et al. 2005, (4) Masciadri et al. 2005, (5) Biller et al. 2007, (6) Kasper et al. 2007, (7) Lafreni\`ere et al. 2007}
\item[\scriptsize{- (*):}] \scriptsize{half have age estimates younger than 200~Myr (see Fig.~1, Lafreni\`ere et al. 2007)}
\end{list}
\end{table}

Deep imaging surveys were performed on other classes of targets:
distant young associations (Taurus, Chamaeleon, Lupus, Upper Sco),
nearby intermediate-age ($0.1-1.0$~Gyr) stars, very nearby stars and
old stars with planets detected by RV. Additional substellar
companions were detected with masses at the edge or inside the
planetary mass regime around the star, DH\,Tau (Itoh et al. 2005), GQ\,Lup
(Neuh\"auser et al. 2005), CHXR\,73 (Luhman et al. 2006), HD230030
(Metchev et al. 2006) and more recently 1RXS J160929.1-210524
(Lafreni\`ere et al. 2008) and CT Cha (Schmidt et al. 2008). The
uncertain fraction of brown dwarf secondaries led various teams
(McCarthy \& Zuckerman 2004; Carson et al. 2005, 2006; Metchev et
al. 2008) to test the extension of the brown desert desert to
intermediate separations. Another purpose was to probe the existence
and the impact of wide massive substellar companions to exoplanetary
systems detected by RV (Patience et al. 2002; Luhman \& Jayawardhana
2002; Chauvin et al. 2006; Mugrauer et al. 2007; Eggenberger et
al. 2007). Only very recently, an important breakthrough has been
achieved with the imaging detection of planetary mass
  companions HR\,8799\,bcd (Marois et al. 2008b), 
  Fomalhaut~b (Kalas et al. 2008) and the candidate $\beta$ Pic\,b (Lagrange et
  al. 2009). Such discoveries offer new attractive perspectives for
current on-going surveys until the arrival of the second generation of
deep imaging instruments like Gemini Planet Imager (GPI; Macintosh et
al. 2006) and VLT/SPHERE (Dohlen et al. 2006).
 
%
% - positive detections: 
%   Physical and chemical properties? Origin?
% - several surveys with null results, different high contrast techniques (coronagraphy, differential imaging, L-band)
%   Statistical analysis of the detection performances to constrain the
%   fraction of BD companion and study the extension of the BD desert
%   at intermdiate sep (McCarthy \& Zuckerman 2004) or to constrain the
%   physical and orbital properties (mass, period, eccentricity
%   districutions) of the existence of a giant planet popualtion arounf
%   those stars (Kasper et al. 2007; Nielsen et al. 2008).

\begin{figure*}[t]
\centering
\includegraphics[width=5.6cm]{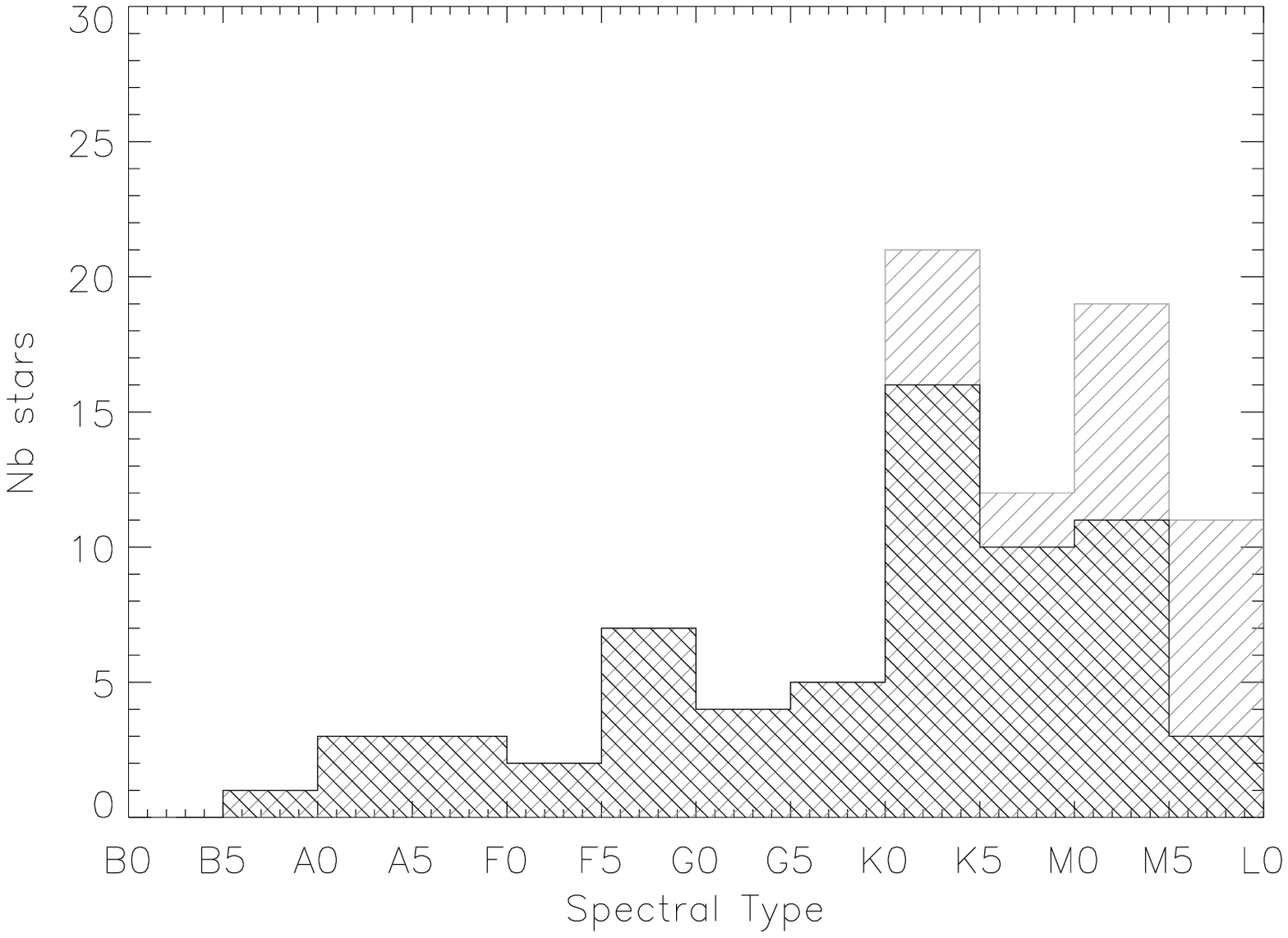}
\includegraphics[width=5.6cm]{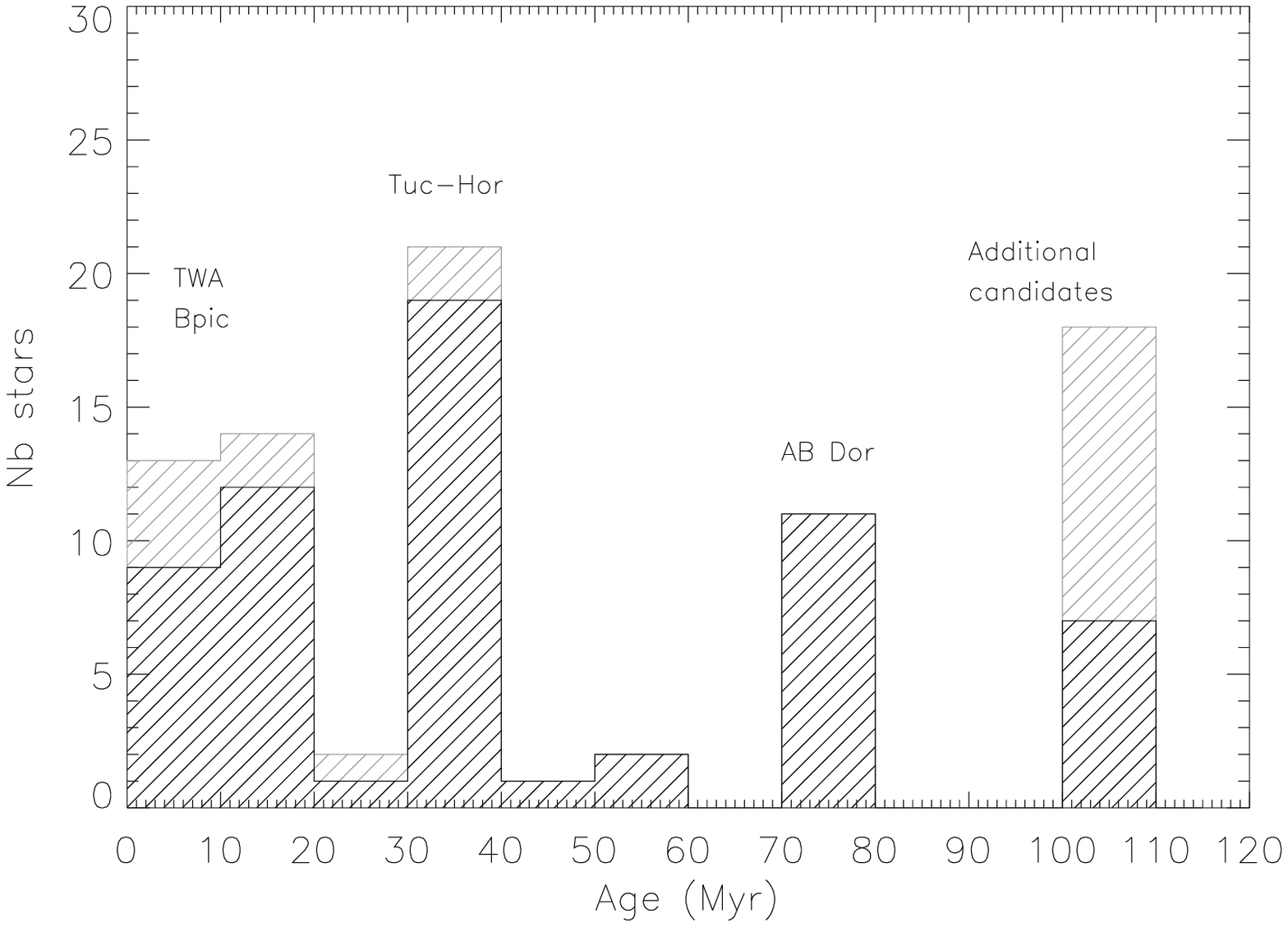}
\includegraphics[width=5.6cm]{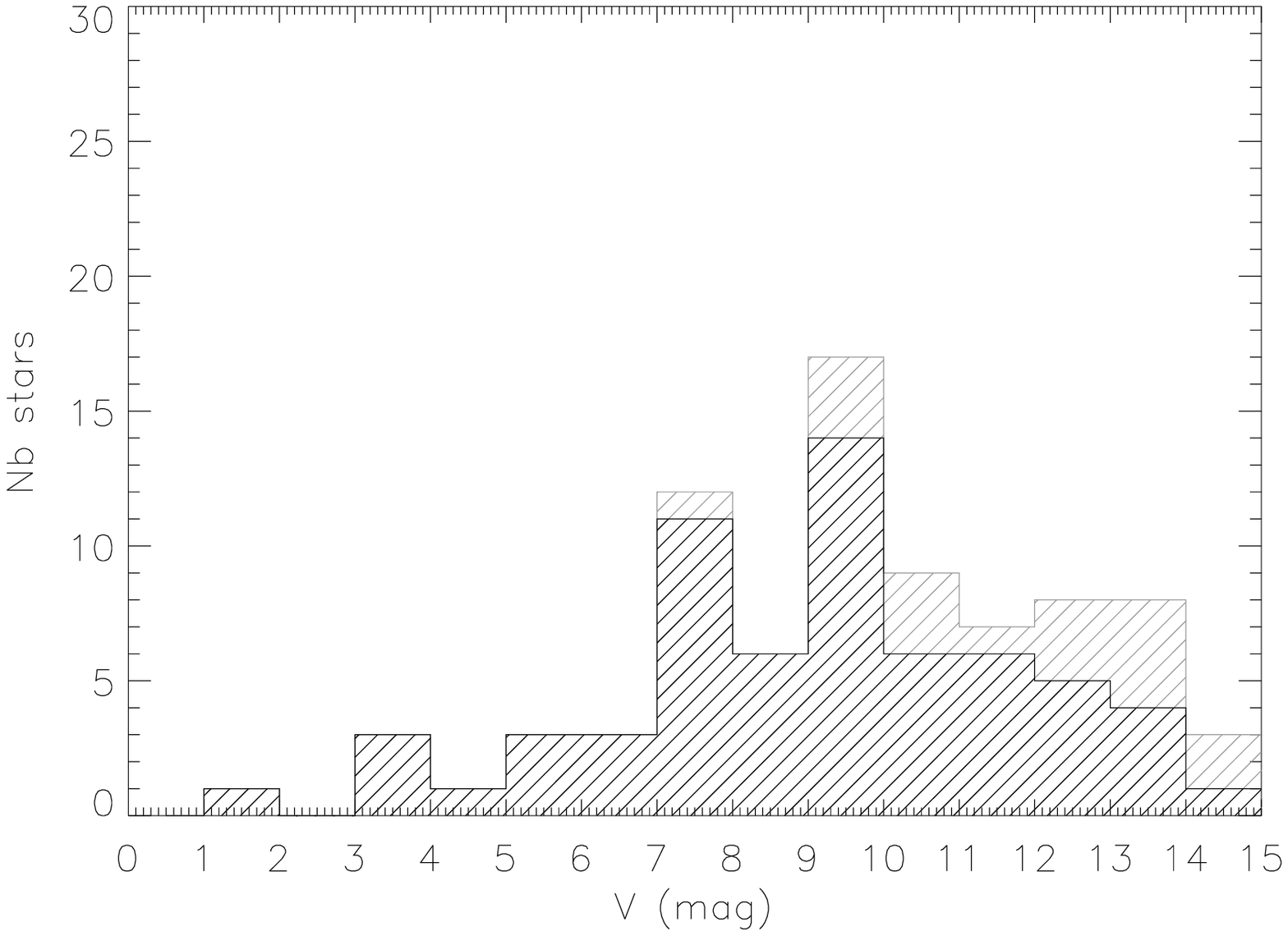}\\
\includegraphics[width=5.6cm]{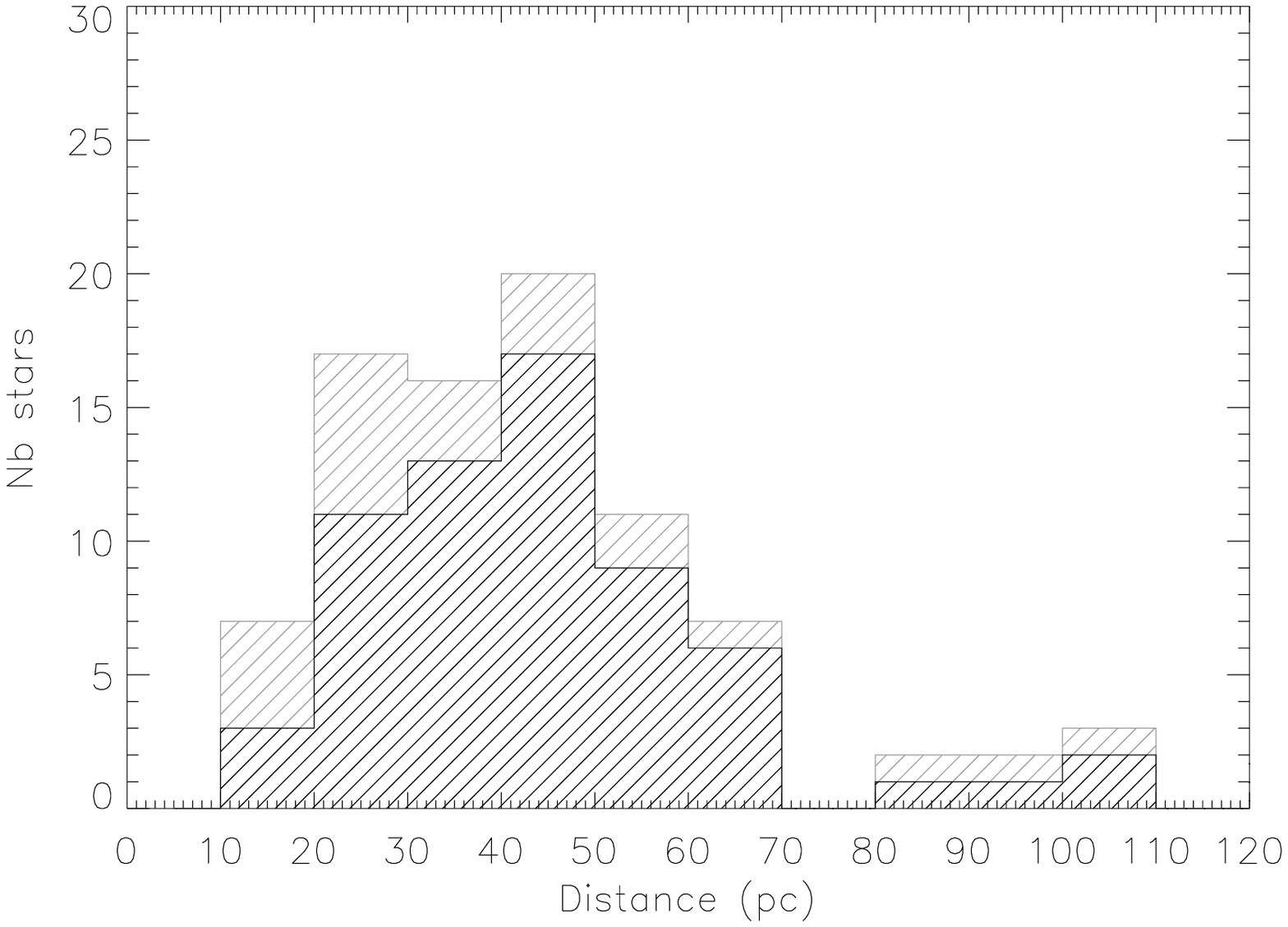}
\includegraphics[width=5.6cm]{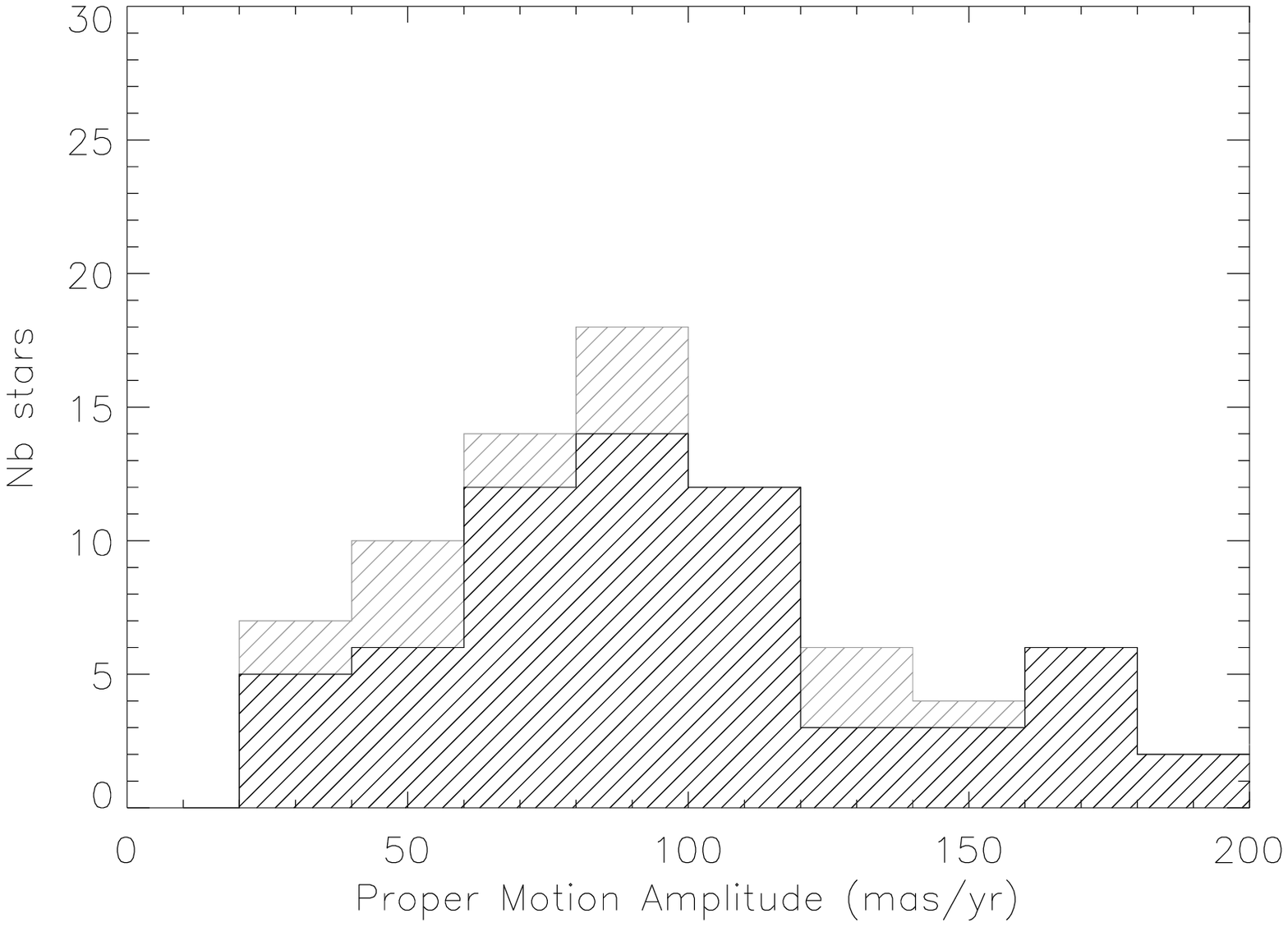}
\includegraphics[width=5.6cm]{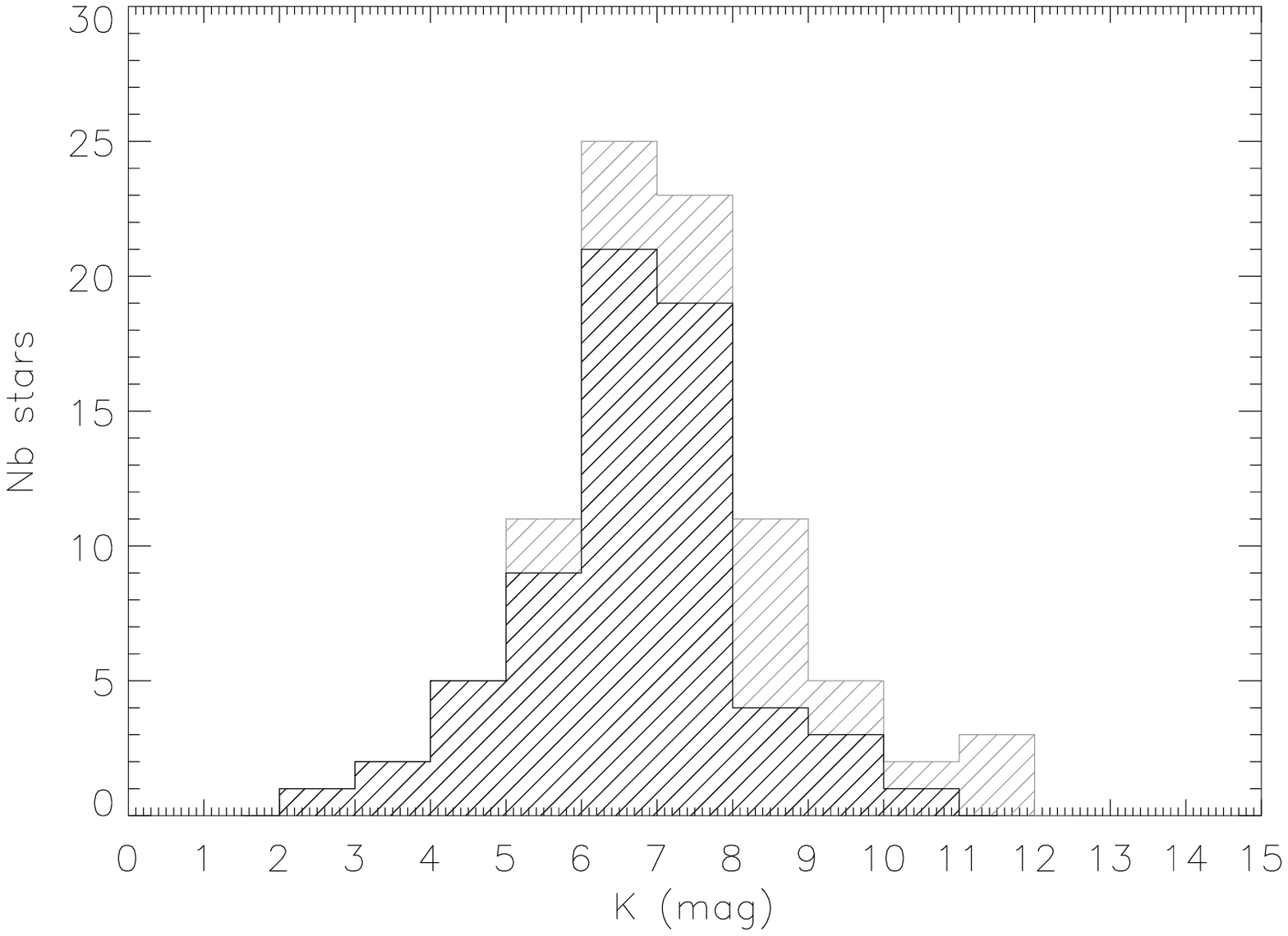}\\
\caption{Histrograms summarizing the main properties of the sample of
young, nearby stars observed with NACO at VLT. \textit{Top-Left:}
Histogram of spectral types for the stars observed in coronagraphic
imaging (\textit{crossed lines}) and in direct imaging (\textit{simple
lines}).  \textit{Top-Middle:} Histogram of ages for members of known
young, nearby associtations (TWA, $\beta$ Pic, Tuc-Hor, AB Dor) and
additional young candidates.  \textit{Top-Right:} Histogram of
$V$-band fluxes. The performances of the AO correction with the NACO
visible-WFS decreases between $12\le V \le 16$. \textit{Bottom-Right:}
Histogram of $K$-band fluxes. The coronagraphic mode is not efficient
anymore for stars fainter than $K\ge9-10$.}
\label{fig:sample_prop}
\end{figure*}

In this paper we report results of a deep coronographic imaging survey
of several young, nearby austral stars, aimed at discovering
substellar companions. In regards to previous works (see Table~1), it
represents one of the largest and deepest surveys obtained so far on this
class of targets. This survey, intitiated in November 2000 with the
ADONIS/SHARPII instrument on a 3.6 m telescope (Chauvin et al. 2003),
was then extended with the VLT/NACO instrument between November 2002
and October 2007. In Section~2, the sample definition and properties
are presented. In Section~3, we describe the characteristics of the
VLT/NACO instrument and the different observing set-up and modes that
we used. The different observing campaigns, the atmospheric conditions
and the observing strategy are detailed in Section~4. The dedicated
data reduction and analysis to clean the science images, to calibrate
our measurments, to derive the relative position and photometry of the
detected sources in the NACO field of view and to estimate the
detection performances are reported in Section~5. We then present the
main results of our survey in Section~6, including the discovery of
new close binary systems and the identification of background
contaminants and comoving companions. In Section~7, we finally
consider the detection sensitivity of our complete survey to
statistically constrain the physical and orbital properties of the
population of giant planets with $20-150$~AU physical
separations.

%
%__________________________________________________________________
\section{Sample Selection}

%
% Subsection Outlines: Sample selection
%
% - Synergy btw identification vs observation
% - Young nearby stars identification
% - Criretia: Youth indicators, distance, Paranal observability, binarity
% - Decomposition in 2 sub-lists
%   1. Deep imaging. Coronagraphy.
%      main properties in table and graphs
%   2. Low-mass primaries (stars and BDs) 
%      main properties in table and graphs
%

The building up of our target sample relied on a synergy between the
exhaustive work of young, nearby stars identification and selection
criteria (age, distance, binarity and observability) to optimize the
detection of close-in planetary mass companions with NACO at
VLT.  Youth
indicators generally rely on the use of photometry and pre-main
sequence isochrones, spectroscopic line (Lithium and H$_\alpha$)
analysis and study of X-ray activity and IR excess (see ZS04).  Association
membership is inferred from coordinates, proper motion, radial
velocity and distance estimation.  Since the beginning of the present
survey, the number of known young, nearby stars more than doubled and
newly identified members were regularly included in our target
sample. Previously known binaries (see
Tables~\ref{table:sample_prop_1} and~\ref{table:sample_prop_2}) with
$1.0-12.0~\!''$ separation were excluded to avoid degrading the NACO
AO and/or coronagraphic detection performances. Our initial complete
sample was composed of 88 stars; 51 are members of young, nearby
comoving groups, 32 are young, nearby stars currently not identified
as members of any currently known association and 5 have been
reclassified by us as older ($>$100~Myr) systems.

%__________________________________________________________________
% Table of young, nearby stars

\begin{table*}[!t]

\caption{Sample of southern young, nearby stars observed during our
VLT/NACO deep imaging survey. In addition to name, coordinates,
galactic latitude (\textit{b}), spectral type, distance and $V$ and
$K$ photometry, the observing filter is given. All sources were
observed in direct imaging, we have therefore indicated the 65 stars
observed in addition in coronagraphy (CI). Finally, the multiplicity
status of the primary and the presence of companion candidates (CCs)
are also reported.  For the multiplicity status we have flagged the
following information: binary (B), triple (T) and quadruple (Q); new
(N) or known/cataloged (K) multiple system; identified visual (VIS),
Hipparcos astrometric (HIP) and spectroscopic (SB) binary system; and
a final flag in case of a confirmed physical (Ph) or comoving (Co)
system, but nothing if only an optical binary.  FS stars are from a
paper by Fuhrmeister \& Schmitt (2003).}

\label{table:sample_prop_1}      
\centering          
\begin{tabular*}{\textwidth}{@{\excs}llllllllllll}     % 7 columns 2
\hline\hline       
                      % To combine 4 columns into a single one 
Name            & $\alpha$        &  $\delta$       & \textit{b}    &    SpT   & \textit{d}  &  Age         & V     & K      &  Mode    & Stellar  &   Note\\
                & [J2000]         & [J2000]         &  (deg)        &          & (pc)        &  (Myr)       & (mag) & (mag)  &   \& Filter & Multiplicity   &  \\
\hline                   
\multicolumn{12}{c}{TWA}\\
\hline 

 TWA22\,AB       & 10 17 26.9  & -53 54 28   &    2  & M5  &  18    &   8  &  13.2  &   7.69   & CI, Ks         & B \tiny{(N/VIS/Ph)} & CCs    \\
 SSSPMJ1102      & 11 02 09.83 & -34 30 35   &   23  & M8  &  65    &   8  &        &   11.88  & Ks            &                     &        \\
 TWA3\,AB        & 11 10 28.8  & -37 32 04   &   21  & M3  &  42    &   8  &  12.1  &   6.77   & CI, H          & B \tiny{(K/VIS/Co)} &        \\
 Twa14           & 11 13 26.3  & -45 23 43   &   14  & M0  &  63    &   8  &  13.8  &   8.50   & CI, Ks         &                     & CCs    \\
 Twa12           & 11 21 05.6  & -38 45 16   &   21  & M2  &  32    &   8  &  13.6  &   8.05   & CI, Ks         &                     & CCs    \\
 2M1139          & 11 39 51.1  & -31 59 21   &   28  & M8  &  49    &   8  &        &  11.50   & Ks             &                     &        \\
 HIP57524        & 11 47 24.6  & -49 53 03   &   11  & G5  & 104    &   8  &  9.1   &   7.51   & CI, H          &                     & CCs    \\
 Twa23           & 12 07 27.4  & -32 47 0    &   30  & M1  &  37    &   8  &  12.7  &   7.75   & CI, H          &                     &        \\
 2M1207          & 12 07 33.4  & -39 32 54   &   23  & M8  &  52    &   8  &        &  11.95   & Ks             & B  \tiny{(N/VIS/Co)}&        \\
 Twa25           & 12 15 30.7  & -39 48 42   &   22  & M5  &  44    &   8  &  11.4  &   7.31   & CI, Ks         &                     &        \\
 HR4796\,A       & 12 36 01.0  & -39 52 10   &   23  & A0  &  67    &   8  &  5.8   &   5.77   & CI, H          & B  \tiny{(K/VIS/Co)}&        \\
 Twa17           & 13 20 45.4  & -46 11 38   &   17  & K5  & 133    &   8  &  12.6  &   9.01   & CI, H          &                     & CCs    \\
\hline                    
\multicolumn{12}{c}{$\beta$ Pictoris}\\
\hline 
 HIP27321        & 05 47 17.0  & -51 03 59   &  -31  & A5  &  20    &  12  &  3.9   &   3.53   & CI, Ks         &                     &        \\
 V343Nor\,B      & 15 38 56.9  & -57 42 18   &  -2   & M4  &  40.0  &  12  &  14.8  &   9.19   & CI, H          & T \tiny{(K/VIS+SB2/Co+Ph)}&  \\
 HD155555\,AB    & 17 17 25.5  & -66 57 03   &  -16  & K1  &  31.4  &  12  &  6.9   &   4.70   & CI, H          & T \tiny{(K/SB2+VIS/Ph+Co)}&CCs\\
 TYC-8742-2065\,AB& 17 48 33.7 & -53 06 43   &  -13  & K0  &  42    &  12  &  9.0   &   6.78   & H             & B \tiny{(K/SB2 and VIS/Ph)}  &  \\
 HIP88399\,A     & 18 03 03.4  & -51 38 56   &  -14  & F5  &  46.9  &  12  &  7.0   &   5.91   & CI, Ks         & B \tiny{(K/VIS/Co)} & CCs    \\
 HIP92024        & 18 45 26.9  & -64 52 16   &  -24  & A7V &  29.2  &  12  &  4.8   &   4.25   & CI, Ks         &                     & CCs    \\
 CD-641208\,AB   & 18 45 37.0  & -64 51 46   &  -24  & K7  &  29.2  &  12  &  9.5   &   6.10   & CI, H          & B \tiny{(N/VIS)}    &        \\
 0ES1847         & 18 50 44.5  & -31 47 47   &  -14  & K5  &  50    &  12  &  10.9  &   7.46   & CI, H          &                     & CCs    \\
 HIP92680        & 18 53 05.8  & -50 10 49   &  -21  & K0V &  49.6  &  12  &  8.4   &   6.37   & CI, Ks         &                     & CCs    \\
 HIP95270        & 19 22 58.9  & -54 32 16   &  -26  & F5  &  50.6  &  12  &  7.0   &   5.91   & CI, H          &                     & CCs    \\
\hline                    
\multicolumn{12}{c}{Tucana-Horologium}\\
\hline 
 HIP1113         & 00 13 53.01 & -74 41 17   &  -42  & G6V &  43.7  &  30  &  8.7   &   6.96   & CI, Ks         &                     &        \\
 HIP1481         & 00 18 26.1  & -63 28 38   &  -59  & F9V &  41.0  &  30  &  8.0   &   6.15   & CI, Ks         &                     & CCs    \\
 CD-7824         & 00 42 20.2  & -77 47 40   &  -40  & K5  &  69    &  30  &  10.4  &   7.53   & CI, H          &                     &        \\
 HIP3556         & 00 45 28.1  & -51 37 33   &  -58  & M1  &  38.5  &  30  & 11.9   &   7.62   & CI, Ks         &                     & CCs    \\
 HIP6485         & 01 23 21.2  & -57 28 50   &  -59  & G6  &  49.3  &  30  &  8.5   &   6.85   & CI, Ks         &                     & CCs    \\
 HIP6856         & 01 28 08.6  & -52 38 19   &  -64  & K1  &  37.1  &  30  &  9.1   &   6.83   & CI, Ks         &                     & CCs    \\
 HD13246\,AB     & 02 07 26.1  & -59 40 45   &  -55  & F8V &  45.0  &  30  &  7.5   &   6.20   & CI, Ks         & B \tiny{(K/SB and VIS/Ph)}&  \\
 GSC08056-00482  & 02 36 51.5  & -52 03 04   &  -58  & M3  &  25    &  30  &  12.1  &   7.50   & CI, Ks         &                     &        \\
 HIP21632\,B     & 04 38 45.6  & -27 02 02   &  -40  & M3V &  54.7  &  30  &  7.5   &  10.41   & CI, Ks$^{*}$   &                     & CCs    \\
 HIP30034        & 06 19 12.9  & -58 03 15   &  -30  & K2  &  45.5  &  30  &  9.1   &   6.98   & CI, H          &                     & CCs    \\
 HIP100751\,AB   & 20 25 38.9  & -56 44 06   &  -35  & B7  &  56    &  30  &  1.9   &   2.48   & CI, Ks         & B \tiny{(K/SB/Ph)}  &        \\
 HIP105404\,ABC  & 21 20 59.8  & -52 28 40   &  -44  & K0V &  46.0  &  30  &  8.9   &   6.57   & CI, Ks         & T \tiny{(K/SB3/Ph)} & CC     \\
 HIP107947       & 21 52 09.7  & -62 03 09   &  -44  & F6  &  45    &  30  &  7.2   &   6.03   & CI, Ks         &                     & CCs    \\
 HIP108195\,ABC  & 21 55 11.4  & -61 53 12   &  -45  & F3  &  47    &  30  &  5.9   &   4.91   & CI, Ks         & T \tiny{(K+N/VIS/Ph+Co)}& CCs   \\
\hline                    
\multicolumn{12}{c}{AB Dor}\\
\hline 
 HIP5191\,A      & 01 06 26.1  & -14 17 47   &  -76  & K1  &  50    &  70  &  9.5   &   7.34   & CI, H          & B \tiny{(K/VIS/Co)} &        \\
 HIP25283        & 05 24 30.2  & -38 58 11   &  -33  & K7  &  18    &  70  &  9.2   &   5.92   & CI, H          & B \tiny{(K/VIS/Co)} &        \\
 ABDor\,BaBb     & 05 28 44.3  & -65 26 46   &  -33  & M3  &  15    &  70  &  13.0  &   7.34   & CI, H          & Q \tiny{(K/VIS/Ph)} &        \\
 HIP26369        & 05 36 55.1  & -47 57 48   &  -32  & K7  &  24    &  70  &  9.8   &   6.61   & CI, H          & B \tiny{(K/VIS/Co)} &        \\
 HIP26373        & 05 36 56.8  & -47 57 53   &  -32  & K0  &  24    &  70  &  7.9   &   5.81   & CI, H          & B \tiny{(K/VIS/Co)} &        \\
 HIP30314        & 06 22 30.9  & -60 13 07   &  -27  & G0V &  23.5  &  70  &  6.5   &   5.04   & CI, Ks         & B \tiny{(K/VIS?)}   & CCs    \\
 GSC08894-00426  & 06 25 55.4  & -60 03 29   &  -27  & M2  &  22    &  70  &  12.7  &   7.21   & CI, Ks         &                     & CCs    \\
 HIP31878        & 06 39 50.0  & -61 28 42   &  -25  & K7  &  21.9  &  70  &  9.7   &   6.50   & CI, Ks         &                     &        \\
 HIP76768\,AB    & 15 40 28.4  & -18 41 45   &   28  & K7  &  43    &  70  &  10.2  &   6.95   & CI, Ks         & B \tiny{(K/VIS/Co)} & CCs    \\
 HIP113579       & 23 00 19.2  & -26 09 13   &  -65  & G1  &  32    &  70  &  7.5   &   5.94   & CI, Ks         &                     & CCs    \\
 HIP118008       & 23 56 10.7  & -39 03 08   &  -77  & K3  &  22.1  &  70  &  8.2   &   5.91   & CI, H          &                     &        \\
\hline                    
\multicolumn{12}{c}{$\eta$ Cha, Near Cha, Columba and Carina}\\
\hline 
 M0838           & 08 38 51.1  & -79 16 13   &  -22  & M5  &  97    &   6  &  16.5  &  10.43   & Ks             &                     &        \\
 HIP58285(TCha)  & 11 57 13.7  & -79 21 32   &  -16  & F5  &  66.4  &  10  &  11.4  &   6.95   & CI, Ks         &                     & CCs    \\
 GSC08047-00232\,A & 01 52 14.6& -52 19 33   &  -62  & K3  &  85    &  30  &  10.9  &   8.41   & CI, Ks         &  B \tiny{(K/VIS/Co)}&        \\
 TYC-9390-0322\,AB& 05 53 29.1 & -81 56 53   &  -29  & K0  &  54    &  30  &  9.1   &   6.94   & H              &  B \tiny{(N/VIS)}   &        \\
\hline                    
\end{tabular*}
\begin{list}{}{}
\item[\scriptsize{- (*):}] \scriptsize{S13 camera used in that case}
\end{list}
\end{table*}

\begin{table*}[t]
\caption{Sample of southern young, nearby stars observed}             
\label{table:sample_prop_2}      
\centering          
\begin{tabular*}{\textwidth}{@{\excs}llllllllllll}     % 7 columns 
\hline\hline       
                      % To combine 4 columns into a single one 
Name            & $\alpha$        &  $\delta$       & \textit{b}    &    SpT   & \textit{d}  &  Age         & V     & K      &  Mode    & Stellar  &   Note\\
                & [J2000]         & [J2000]         &  (deg)        &          & (pc)        &  (Myr)       & (mag) & (mag)  &   \& Filter & Multiplicity   &  \\
\hline                    
\multicolumn{12}{c}{Additional young candidates}\\
\hline 
 BTR99\,AB       & 01 23 17.0  & -79 41 32   &  -37  & K0  & 103    &  10  &  10.1  &   7.07   & CI, H          &  B \tiny{(N/VIS)}   &        \\
 CD-53386\,AB    & 02 01 53.7  & -52 34 53   &  -61  & K3  & 120    &  30  & 11.0   &   8.60   & H              &  B \tiny{(N/VIS)}   &        \\
 FS75            & 02 04 53.2  & -53 46 16   &  -60  & M4  &  30    & 100  &  15.0  &   9.6    & Ks             &                     &     \\
 FS84            & 02 22 44.2  & -60 22 47   &  -53  & M4  &  20    & 100  &  13.7  &   8.2    & Ks             &                     &        \\
 GSC08862-00019  & 02 58 04.6  & -62 41 15   &  -49  & K4  & 138    &  20  & 11.7   &   8.91   & CI, Ks          &                    & CCs    \\
 TYC6461-1120\,A & 04 00 03.7  & -29 02 16   &  -48  & K0  &  62    &  40  &  9.6   &   7.15   & CI, Ks          & B \tiny{(N/VIS/Co)}& CCs    \\
 HIP28474\,AB    & 06 00 41.3  & -44 53 50   &  -27  & G8  &  53.7  & 100  &  9.1   &   7.32   & CI, H           & B \tiny{(N/VIS)}   &        \\
 FS388\,ABC      & 06 43 45.3  & -64 24 39   &  -25  & M4  &  22    & 100  &  14.0  &   8.4    & Ks             & T \tiny{(N/VIS)}   &        \\
 FS465\,AB       & 08 17 39.4  & -82 43 30   &  -24  & M4  &  10    & 100  &  12.6  &   6.6    & Ks             & B \tiny{(N/VIS)}   &     \\
 HIP41307        & 08 25 39.6  & -03 54 23   &   18  & A0  &  38    & 100  &  3.9   &   4.08   & CI, Ks          &                    &        \\
 FS485           & 08 47 22.6  & -49 59 57   &   -4  & M2  &  33    & 100  &  12.0  &   7.71   & Ks             &                    &        \\
 FS488\,AB       & 08 54 02.4  & -30 51 36   &    9  & M5  &  15    & 100  &  13.4  &   8.10   & Ks             & B \tiny{(N/VIS)}   &        \\
 HIP51386        & 10 29 42.2  & +01 29 28   &   47  & F5  &  31.5  &  50  &  6.9   &   5.52   & CI, Ks          &                    & CCs    \\
 FS588           & 11 20 06.1  & -10 29 47   &   46  & M3  &  20    & 100  &  12.1  &   7.0    & Ks             &                    &        \\
 HIP59315        & 12 10 06.4  & -49 10 50   &   13  & G5  &  37.8  & 100  &  8.2   &   6.50   & CI, H           &                    & CCs    \\
 CD-497027       & 12 21 55.6  & -49 46 12   &   13  & K0  &  89    &  20  & 10.1   &   8.01   & Ks             &                    &        \\
 HIP61468        & 12 35 45.5  & -41 01 19   &   21  & A7  &  34.6  & 100  &  5.1   &   4.57   & CI, H           &                    &        \\
 TYC-8992-0605   & 12 36 38.9  & -63 44 43   &    0  & K3  &  50    &  10  &  9.9   &   7.37   & CI, H           &                    & CCs    \\
 TYC-09012-1005  & 13 44 42.6  & -63 47 49   &   -1  & K5  &  95    &  10  & 11.0   &   7.74   & CI, H           &                    & CCs    \\
 TYC-7818-0504\,AB& 14 30 13.5 & -43 50 09   &   16  & K5  & 100    &  10  & 10.4   &   7.64   & H              & B \tiny{(N/VIS)}   &        \\
 HIP74405        & 15 12 23.4  & -75 15 15   &  -15  & K0  &  50.2  & 100  &  9.4   &   7.38   & CI, H           &                    &        \\
 TYC-7846-1538   & 15 53 27.3  & -42 16 02   &    9  & G1  &  48    &  30  &  7.9   &   6.34   & CI, H           &                    & CCs    \\
 HIP80448\,ABC   & 16 25 17.5  & -49 08 52   &    0  & K1  &  45.5  & 100  &  7.1   &   5.70   & H              & T \tiny{(K/SB+VIS/Ph+Co)}&     \\
 HIP84642\,AB    & 17 18 14.7  & -60 27 27   &  -13  & K0  &  54.6  &  40  &  9.5   &   7.53   & CI, Ks          & B \tiny{(N/VIS)}   & CCs    \\
 FS903           & 17 37 46.5  & -13 14 47   &    9  & K7  &  45    & 100  &  10.2  &   6.835  & CI, Ks          &                    & CCs    \\
 FS979\,AB       & 18 35 20.8  & -31 23 24   &  -11  & M5  &  18    & 100  &  13.1  &   7.8    & Ks             & B \tiny{(N/VIS)}   & CCs    \\
 FS1017          & 19 19 20.2  & -01 33 54   &   -6  & M5  &  25    & 100  &  16.6  &   9.667  & Ks             &                    & CCs    \\
 FS1035          & 19 42 12.8  & -20 45 48   &  -20  & M5  &  20    & 100  &  14.4  &   8.756  & Ks             &                    & CCs    \\
 HIP98495        & 20 00 35.5  & -72 54 37   &  -31  & A0  &  33.3  &  50  &  3.9   &   3.80   & CI, H           &                    &        \\
 HIP102626       & 20 47 45.0  & -36 35 40   &  -38  & K0  &  44.4  &  30  &  9.4   &   6.79   & CI, H           & B \tiny{(K/HIP?)}  &        \\ 
 FS1136\,AB      & 21 49 06.2  & -64 12 55   &  -43  & M5  &  25    & 100  &  15.5  &   9.5    & CI, Ks          & B \tiny{(N/VIS)}   &        \\
 FS1174          & 22 44 08.0  & -54 13 20   &  -54  & M4  &  30    & 100  &  13.4  &   8.5    & Ks             &                    & CCs    \\
\hline                    
\multicolumn{12}{c}{Reclassified as older systems}\\
\hline 
 HIP7805         & 01 40 24.1  & -60 59 57   &  -55  & F2  &  67    &$\ge 100$&7.7  &   6.63   & CI, H           &                    &        \\
 HIP69562\,ABC   & 14 14 21.3  & -15 21 21   &   42  & K5V &  26.5  &$\ge 100$&10.5 &   6.60   & Ks             & T \tiny{(N/VIS)}   &        \\
 HIP76107        & 15 32 36.7  & -52 21 21   &    3  & M0  &  30.6  &$\ge 100$&11.0 &   7.60   & CI, Ks          & B \tiny{(K/HIP?)}  & CCs    \\
 HIP96334        & 19 35 09.7  & -69 58 32   &  -29  & G1V &  35.4  &$\ge 100$&7.9  &   6.30   & CI, Ks          &                    & CCs    \\
 HIP107705\,AB   & 21 49 05.8  & -72 06 09   &  -39  & M0  &  16.1  &200      &9.8  &   5.65   & Ks             & B \tiny{(N/VIS)}   &        \\
 \hline                  
\end{tabular*}
\end{table*}

For stars not in a known moving group (Table~3), based on existing
data we employed as many of the techniques for age dating as possible
(see, e.g., Section 3 in ZS04).  The principal diagnostics were
Lithium abundance, Galactic space motion UVW, and fractional X-ray
luminosity (Figs. 3, 6 and 4, respectively in ZS04).  With the
possible exception of a few of the FS stars (see following paragraph),
all Table~3 stars with ages 100 Myr or less have UVW in or near the
"good UVW box" in Fig. 6 of ZS04.  With the exception of the A-type
stars (unknown Lithium abundances), all Table~3 stars have Lithium
abundances (we have measured) consistent with the ages we list and
their spectral type (as per Fig. 3 in ZS04).  With the
exception of the A-type stars, X-ray fluxes are consistent with Fig.~4
in ZS04 for the indicated ages. Age uncertainties for
non-FS stars in Table~3 are typically $±50\%$ of the tabulated age (i.e.,
$30\pm15$~Myr, $100\pm50$~Myr).  The ages of the two A-type stars are
based on UVW and location on a young star HR diagram.

When their radial velocity is known (based on our echelle spectra)
then the FS stars usually have a "good UVW".  In all cases they are
strong X-ray emitters and also have H alpha in emission, usually
strongly.  Lithium is usually not detected in the FS stars, or
occasionally weakly.  Because the data sets for these stars are
sometimes incomplete (e.g., radial velocity not measured) and because
fractional X-ray luminosity and UVW are imprecise measures of age, we
have assigned an age of 100 Myr to all observed FS stars.  Perhaps a
few FS stars have ages older than 100 Myr (FS 588 being the most
likely of these).  But, similarly, some are likely younger than 100
Myr.  By assuming an overall uniform age of 100 Myr for the sample of
FS stars, we are probably somewhat overestimating their mean age. The age determination of the ensemble of FS stars is likely to be accurate to within about a factor 2 in general, although the age of some FS starscould well lie outside of this range.

The sample properties are summarized in
Tables~\ref{table:sample_prop_1} and~\ref{table:sample_prop_2} and
illustrated in Fig.~\ref{fig:sample_prop}. $93\%$ of the selected
stars are younger than about $100$~Myr and $94\%$ closer than
100~pc. The spectral types cover the sequence from B to M spectral
types with $19\%$ BAF stars, $48\%$ GK stars and $33\%$ M dwarfs.

%
%__________________________________________________________________
\section{Observations}

%
%__________________________________________________________________
\subsection{Telescope and instrument}

NACO\footnote{http://www.eso.org/instruments/naos/} is the first
Adaptive Optics instrument that was mounted at the ESO Paranal
Observatory near the end of 2001 (Rousset et
al. 2002).  NACO provides diffraction limited images in the near
infrared (nIR). The observing camera CONICA (Lenzen et al. 2002) is
equipped with a $1024\times1024$ pixel Aladdin InSb array. NACO offers
a Shack-Hartmann visible wavefront sensor and a nIR wavefront
sensor for red cool (M5 or later spectral type) sources. nIR wavefront
sensing was only used on 8\% of our sample. Note that in May 2004, the
CONICA detector was changed and the latter detector was more efficient
thanks to an improved dynamic, a lower readout noise and cleaner
arrays. Among NACO's numerous observing modes, only the direct and
coronagraphic imaging modes were used. The two occulting masks offered
for Lyot-coronagraphy have a diameter of $\oslash=0.7~\!''$.  and
$\oslash=1.4~\!''$.  According to the atmospheric conditions, we used
the broad band filters $H$ and $K_s$, the narrow band filters, NB1.64,
NB1.75 and Br$\gamma$
 \footnote{see filters description: http://www.eso.org/instruments/naco/inst

/filters.html} and a neutral density filter (providing a transmissivity
factor of 0.014). In order to correctly sample the NACO PSF (better
than Nyquist), the S13 and S27 objectives were used, offering mean
plate scales of $13.25$ and $27.01$~mas per pixel and fields of view
of $14~\!''  \times14~\!''$ and $28~\!''  \times28~\!''$ respectively.

%________________________________
%\subsection{Observing campaigns}

%
% Subsection Outlines
%
% - Observations description
% - Observing campaigns
% - observing modes
%

Our deep imaging survey was initiated during guaranteed time
observations shared between different scientific programs and
scheduled between November 2002 and September 2003. The survey was
extended using open time observations between March 2004 and June
2007. The open time observations were shared between classical visitor
mode and remote service mode as offered by ESO at the Paranal
Observatory.  The observing programs, observing modes, the number of
nights allocated, the time loss due to technical or weather reasons
and the number of targets observed per run are reported in
Table~\ref{tab:obs}.

%________________________________
\subsection{Image quality}

%
% Subsection Outlines
%
% - Atmospheric conditions: tau0, seeing, airmass
% - Sr corrections
%

\begin{table}[t]

\caption{Summary of the different observing campaigns of our survey.
For each campaign, we report the ESO programme numbers, the
observation type, Guaranteed Time (GTO) or Open Time (OT), if obtained
in visitor (Vis) or service (Ser) modes, the starting nights of
observation, the number of nights allocated and the time
loss. Finally, the number of visits, corresponging to the number of
observing sequences executed on new and follow-up targets, is given.}

\label{tab:obs}      
\begin{tabular*}{\columnwidth}{@{\excs}llllll}     % 7 columns 
\hline\hline       
ESO Program             & Mode      &  Start. Night   & Night        & Loss       & Visits \\
                        &           &  (UT-date)      & (Nb)         & (\%)       & (Nb) \\
\hline
070.C-0565(A)           & GTO-Vis   &  26-11-2002     & 1            &            & 6            \\ 
070.D-0271(B)           & GTO-Vis   &  16-03-2003     & 1.5          &            & 8            \\  
071.C-0507(A)           & GTO-Vis   &  07-06-2003     & 0.5          & 50         & 2            \\  
071.C-0462(A)           & GTO-Vis   &  07-09-2003     & 0.5          & 50         & 2            \\  
072.C-0644(A)           & OT-Vis    &  05-03-2004     & 1            & 100        & 0            \\  
072.C-0644(B)           & OT-Vis    &  05-03-2004     & 1            & 0          & 9            \\  
073.C-0469(A)           & OT-Vis    &  27-04-2004     & 1            & 0          & 16           \\  
073.C-0469(B)           & OT-Vis    &  25-09-2004     & 1            & 0          & 12           \\  
075.C-0521(A)           & OT-Vis    &  06-05-2005     & 1            & 0          & 15           \\  
075.C-0521(B)           & OT-Vis    &  19-08-2005     & 0.5          & 0          & 10           \\  
076.C-0554(A)           & OT-Vis    &  08-01-2006     & 1            & 0          & 13           \\  
076.C-0554(B)           & OT-Vis    &  26-02-2006     & 1            & 30         & 8            \\  
078.C-0494(A)           & OT-Ser    &  2006/2007      & 0.7          & 30         & 7            \\
079.C-0908(A)           & OT-Ser    &  2007           & 1            & 0          & 10           \\
\hline                  
Total                   & -         & -               & 11.7         &            & 127          \\ 
\hline                  
\end{tabular*}
\end{table}

\begin{figure}[t]
\centering
\includegraphics[width=\columnwidth]{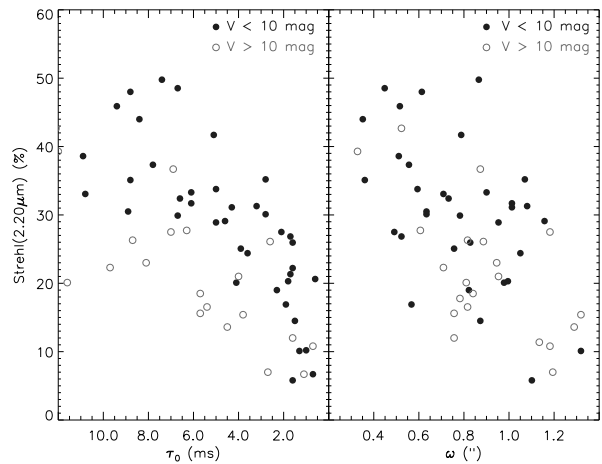}

\caption{VLT/NACO adaptive optics system performances. Strehl ratio at
  2.20~$\mu$m as a function of the correlation time $\tau_0$ and the
  seeing $\omega$ of the atmospheric turbulence for two regimes of
  $V$-band magnitude of the primary star (AO reference target). Only
  the targets observed with the visible WFS are plotted. Close binaries have 
  also been rejected. The results
  demonstrate the good behavior of NACO over a wide range of stellar
  magnitudes and under different turbulent conditions. A clear
  degradation of the performances is seen for decreasing $\tau_0$, increasing
  $\omega$ and fainter visible ($V\ge10$) targets. A clear drop is
  seen for $\tau_0$ faster than 2~ms, the limit of the NACO wavefront sensor sampling frequency.}

\label{fig:obs_conditions}
\end{figure}

For ground-based telescopes, atmospheric conditions have always been
critical to ensure astronomical observations of good quality. Although
AO instruments aim at compensating the distorsion induced by
atmospheric turbulence, the correction quality (generally measured by
the \textit{strehl ratio} and \textit{Full Width Half Maximum (FWHM)} 
parameters) is still
related to the turbulence speed and strength.  For bright targets, the
NACO AO system can correct for turbulence with a coherent time
($\tau_0$) longer than $2$~ms. For faster ($\tau_0\le2$~ms)
turbulence, the system is always late and the image quality and the
precision of astrometric and photometric measurements are consequently
degraded. During our NACO observing runs, the averaged $\tau_0$ was
about 5~ms and larger than 2~ms 80\% of the time. The average seeing
conditions over all runs was equal to 0.8~$\!''$ (which 
happens to be the median seeing value measured in
Paranal over the last
decade\footnote{http://www.eso.org/gen-fac/pubs/astclim/paranal/seeing/adaptive-optics/statfwhm.html}). Fig.~\ref{fig:obs_conditions}
shows the (\textit{strehl ratio}) performances of the NACO AO system
with the visible wavefront sensor as a function of 
the correlation time of the atmosphere $\tau_0$, the seeing and the primary visible
magnitude. As expected, the 
degradation of the performances 
is seen with a decrease of $\tau_0$, the coherent length ($r_0$, inversely proportional to the seeing) and the 
primary flux. Still, the results clearly demonstrate the good NACO
performances and capabilities over a wide range of observing conditions.

%________________________________
\subsection{Observing strategy}

%
% Subsection Outlines
%
% - continuity of ADONIS survey
% - coronagraphy to enhance detection perfomances
% - dedicated to astrometry
% - description of the observing sequence and observing strategy
% - typical integration time
% - strategy in case of posiitive detections
% 

The VLT/NACO survey was conducted as a continuation of our earlier
coronographic survey with the ADONIS/SHARPII instrument at the ESO 3.6
m telescope at La Silla Observatory (Chauvin et al. 2003). A similar
observing strategy was adopted to optimize the detection of faint
close substellar companions. Most of our stars are relatively
bright ($K_s\le10$) in nIR. To improve our detection performances, we
have opted for the use of Lyot coronography. High contrast imaging
techniques, such as Lyot and phase mask coronagraphy, $L$-band
saturated imaging and simultaneous differential imaging, enable
achievement of contrasts of $10^{-5}$ to $10^{-6}$. Their main
differences are inherent in the nature of the substellar companions
searched and the domain of separations explored. Broad-band nIR Lyot
coronagraphy and thermal ($L'$-band or 4~$\mu$m) saturated imaging are
among the most sensitive techniques at typical separations between 1.0 to
$10.0~\!''$. The contrast performances are currently mandatory to
access the planetary mass regime when searching for faint close
companions.

To measure precisely the positions of the faint sources detected in
the coronagraphic field relative to the primary star, a dedicated
observing block was executed. This block was composed of three
successive observing sequences and lasted in total $\sim45$~min
(including pointing). After the centering of the star behind the
coronagraphic mask, a deep coronagraphic observing sequence on source
was started. Several exposures of less than one minute each were
accumulated to monitor the star centering and the AO correction
stability. An effective exposure time of 300~sec was generally spent
on target.  During the second sequence, a neutral density or a narrow
band filter was inserted and the occulting mask and Lyot stop
removed. The goal was to precisely measure the star position behind
the coronagraphic mask (once corrected for filter shifts). An
effective exposure time of 60~sec was spent on source. Counts were
adjusted to stay within the $1\%$ linearity range of the detector. The
image is also used to estimate the quality of the AO
correction. Finally, the last sequence was the coronagraphic sky. This
measure was obtained $\sim45~\!''$ from the star using a jittering
pattern of several offset positions to avoid any contaminants in the
final median sky. In case of positive detections, whenever possible,
the companion candidates (CCs) were re-observed to check whether a
faint object shared common proper motion with the primary
star. Depending on each object's proper motion (see
Fig.~\ref{fig:sample_prop}), the timespan between successive epochs
was about 1-2 years. When comoving companions were identified, images
were recorded with addditional nIR filters to directly compare the
spectral energy distribution with that predicted by (sub)stellar
evolutionary models.

%
%__________________________________________________________________
\section{Data reduction and analysis}

%
% Data Reduction and Analaysis
%
% - Reduction, cosmetic
% - PSF subtraction
% - Photometry and Astrometry estimation
% - Detection Limit estimation
%

%________________________________
\subsection{Cosmetic and image processing}

\begin{figure}[t]
\centering
\includegraphics[width=\columnwidth]{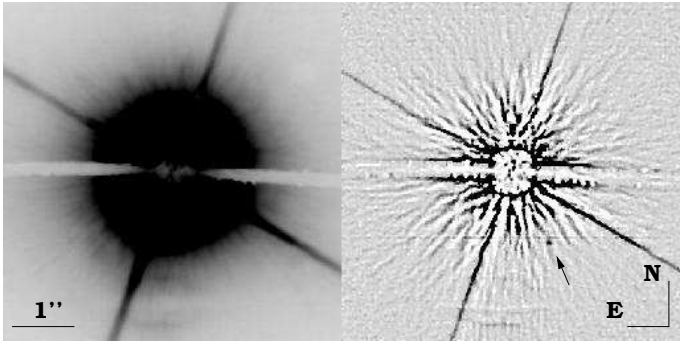}

\caption{\textit{Left}: VLT/NACO corongraphic image of HIP\,95270
obtained in $H$-band with the S13 camera. The small ($\oslash =
0.7~\!''$) coronagraphic mask was used. \textit{Right}: Coronagraphic
image after high-pass filtering. A kernel of $3 \times FWHM$ is used
to remove the low spatial frequencies of the coronagraphic PSF
wings. A fake $\Delta H = 12$ companion has been inserted at
1.2$~\!''$ from the star to test the detection performances. Minimum
and maximum thresholds of the filtered image were divided by a factor
15 to show the fake companion and the PSF residuals.}

\label{fig:image_corono}
\end{figure}

Classical cosmetic reduction including bad pixels removal,
flat-fielding, sky substraction and shift-and-add, was made with the
\textit{Eclipse}\footnote{http://www.eso.org/projects/aot/eclipse/}
reduction software developed by Devillar (1997) for both direct and
coronagraphic imaging observations. Median filtering by a kernel of $3
\times 3$ pixels was applied to correct for remaining hot pixels. To
remove the central part of the PSF in our reduced coronographic
images, two methods were applied. The first method considered
different angular sectors uncontaminated by the diffraction spikes and
by the coronographic mask support. For each sector, the PSF azimuthal
average is calculated, circularised and subtracted from the
coronagraphic image. The alternative method was to apply directly a
high-pass filter with a kernel of $3 \times FWHM$ (assuming the
theoretical \textit{FWHM} at each observing set-up). As a
result, low spatial frequencies, including the coronagraphic PSF
wings, were removed from the reduced image. Finally, each resulting
image was inspected by eye for the CCs identification.
Fig.~\ref{fig:image_corono} is an illustration of the data processing
applied to the coronagraphic images of HIP\,95270, in the case of the
second method.

%________________________________
\subsection{Astrometric calibration}

The astrometric calibration of high angular resolution images as
provided by NACO is not a simple task. As NACO is not a
multi-conjugated AO system, the diffraction limited images have a
small FoV limited by the anisoplanetism angle. Therefore, classical
high-precision astrometric techniques over crowded fields of thousands
of stars cannot be transposed. In addition, ESO does not currently
provide any detector distorsion map. For this reason, astrometric
calibrators were observed within a week for each observing run (in
visitor and service mode) to determine a mean platescale and the true
north orientation. Our primary astrometric calibrator was the
$\Theta_1$ Ori C field observed with HST by McCaughrean \& Stauffer
(1994). The same set of stars (TCC058, 057, 054, 034 and 026) were
observed with the same observing set-up ($K_s$ with S27 and $H$ with
S13) to avoid introduction of systematic errors.  When not observable,
we used as secondary calibrator the astrometric binary IDS21506S5133
(van Dessel \& Sinachopoulos 1993), yearly recalibrated with the
$\Theta_1$ Ori C field. The mean orientation of true north and the
mean platescale of the S13 and S27 cameras are reported in
Table~\ref{tab:calastro}.

\begin{table}[t]
\caption{Mean plate scale and true north orientation for each observing
run. The astrometric field $\Theta_1$ Ori C field and the astrometric
binary IDS21506S5133 (van Dessel \& Sinachopoulos 1993) were the calibrators.}
\label{tab:calastro}      
\begin{tabular*}{\columnwidth}{@{\excs}lllll}     % 7 columns 
\hline\hline       
ESO Program     & UT Date                 &       Obj.       & Platescale        & True north     \\
                &                         &                  & (mas)             & (deg)              \\
\hline
070.C-0565      & 21-11-2002              & S13              & $13.24\pm0.05$    & $-0.05\pm0.10$   \\
                & 21-11-2002              & S27              & $27.01\pm0.05$    & $0.08\pm0.18$    \\
070.D-0271      & 16-03-2003              & S13              & $13.21\pm0.11$    & $-0.05\pm0.10$   \\
071.C-0507      & 29-05-2003              & S13              & $13.24\pm0.05$    & $-0.10\pm0.10$   \\
                & 03-06-2003              & S27              & $27.01\pm0.05$    & $0.01\pm0.19$    \\
071.C-0507      & 07-09-2003              & S13              & $13.24\pm0.05$    & $0.05\pm0.10$    \\
072.C-0644      & 05-03-2004              & S13              & $13.24\pm0.05$    & $0.04\pm0.10$    \\
                & 05-03-2004              & S27              & $27.01\pm0.05$    & $-0.18\pm0.20$   \\
073.C-0469      & 27-04-2004              & S27              & $27.01\pm0.05$    & $0.08\pm0.20$    \\
073.C-0469      & 22-09-2004              & S13              & $13.25\pm0.05$    & $0.20\pm0.10$    \\
                & 22-09-2004              & S27              & $27.01\pm0.05$    & $0.0\pm0.19$     \\
075.C-0521      & 19-08-2005              & S13              & $13.25\pm0.06$    & $-0.02\pm0.10$   \\
                & 19-08-2005              & S27              & $27.01\pm0.06$    & $-0.07\pm0.11$   \\
076.C-0654      & 08-01-2006              & S13              & $13.25\pm0.06$    & $0.18\pm0.10$    \\
                & 08-01-2006              & S27              & $27.02\pm0.06$    & $0.12\pm0.13$    \\
076.C-0654      & 28-02-2006              & S13              & $13.25\pm0.06$    & $0.19\pm0.10$    \\
                & 28-02-2006              & S27              & $27.02\pm0.05$    & $0.13\pm0.14$    \\
078.C-0494      & 24-10-2006              & S13              & $13.26\pm0.07$    & $-0.19\pm0.23$   \\
                & 23-12-2006              & S13              & $13.26\pm0.08$    & $-0.23\pm0.15$   \\
                & 22-10-2006              & S27              & $27.01\pm0.03$    & $-0.30\pm0.16$   \\
                & 25-12-2006              & S27              & $27.01\pm0.04$    & $-0.20\pm0.18$   \\
079.C-0908      & 18-07-2007              & S27              & $27.01\pm0.05$    & $-0.06\pm0.15$   \\
\hline                  
\end{tabular*}
\end{table}

%________________________________
\subsection{Companion candidate characterization}

%
% Subsection Outlines
%
% - DI and COR analysis
% - Case of positive detection
% - Method to test if comoving or not
%

For direct imaging, the relative photometry and astrometry of 
visual binaries were obtained using the classical deconvolution
algorithm of V\'eran \& Rigaut (1998). This algorithm is particularly
adapted for stellar field analysis. Several PSF references were used
to measure the influence of the AO correction. They were selected to
optimize a set of observing criteria relative to the target
observation (observing time, airmass, spectral type and $V$ or
$K$-band flux according to the wavefront sensor).

\begin{figure*}[t]
\centering
\includegraphics[width=\columnwidth]{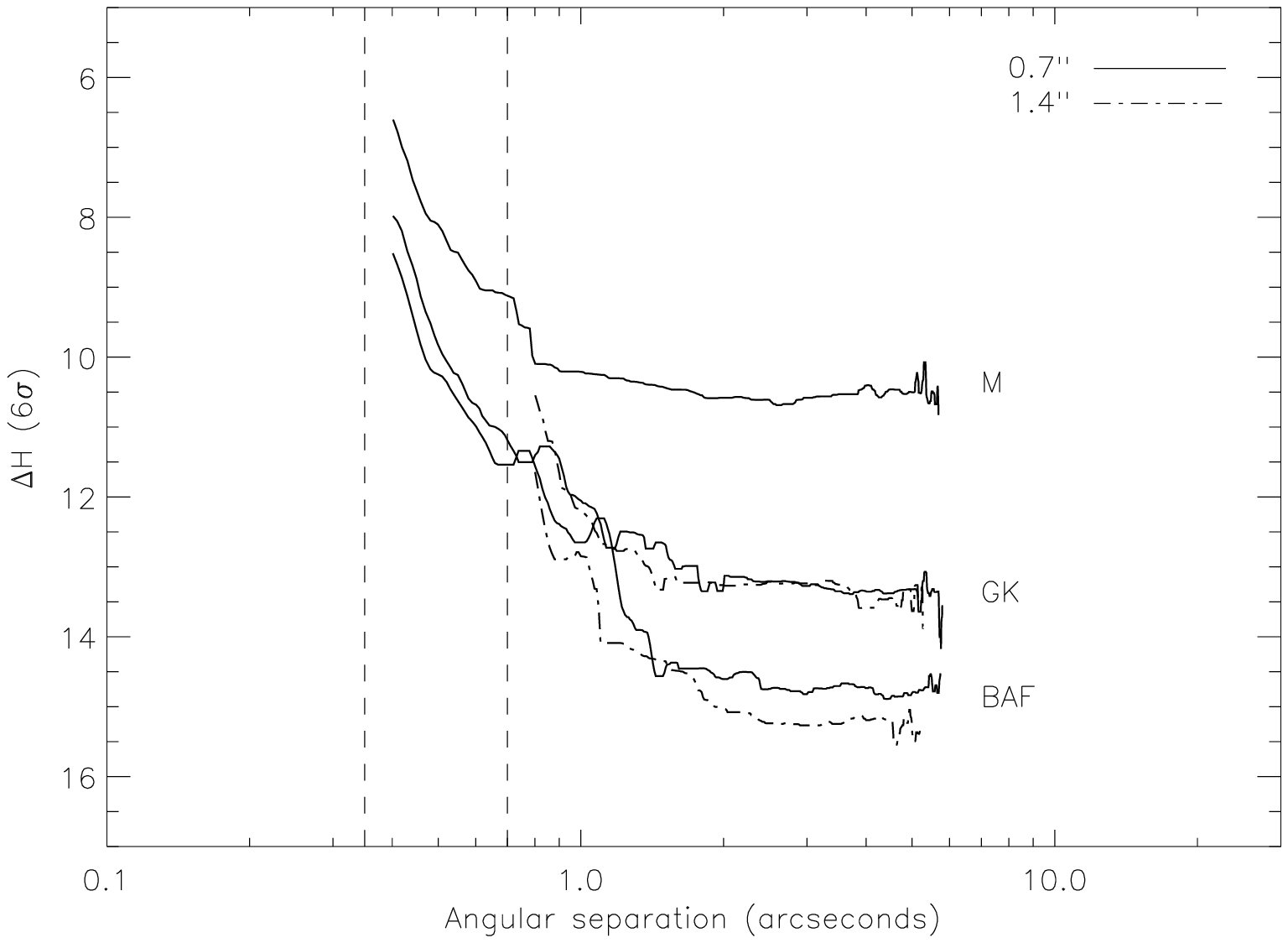}
\includegraphics[width=\columnwidth]{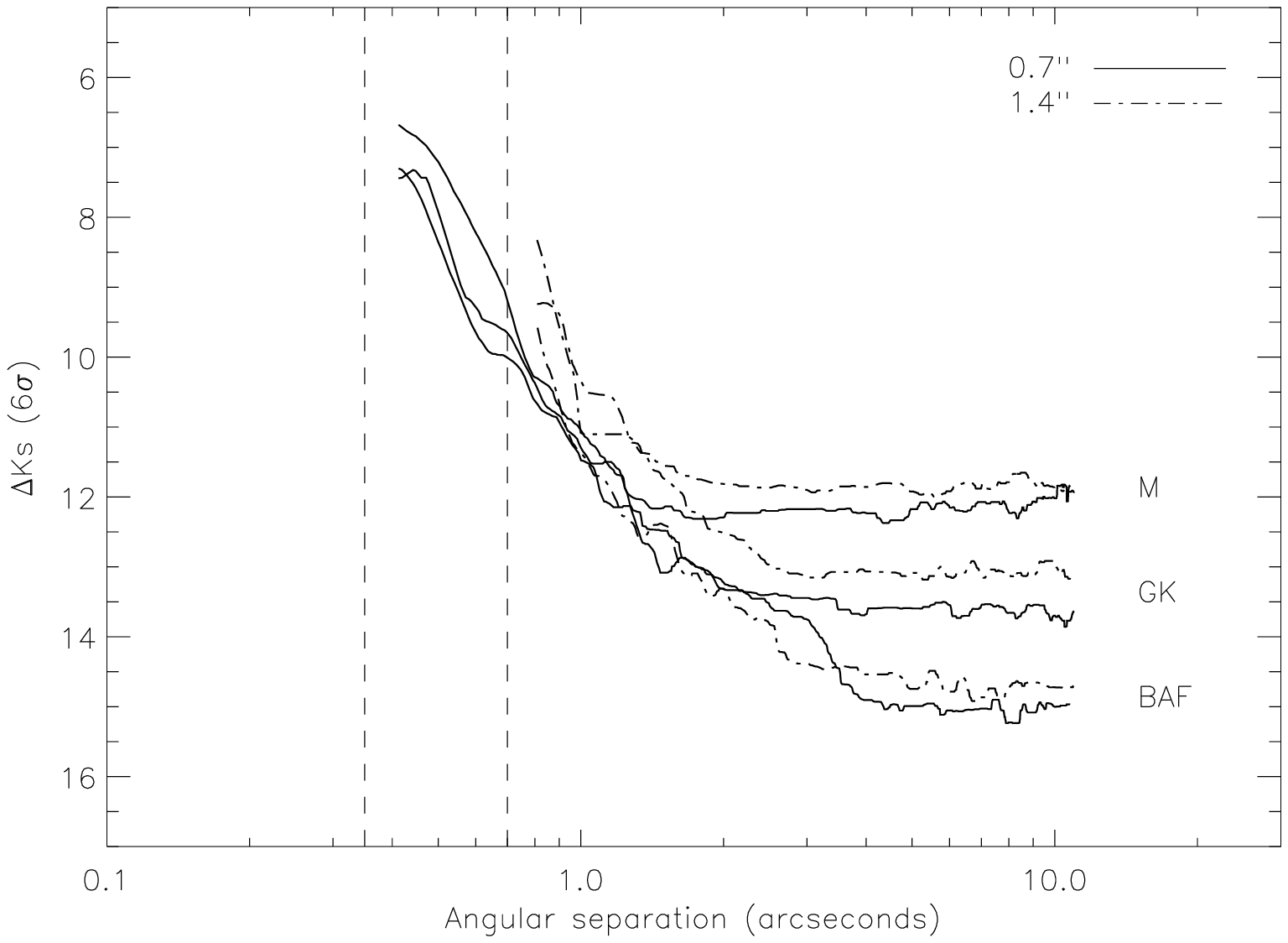}

\caption{\textit{Left:} VLT/NACO coronagraphic detection limits in
H-band (combined with the S13 camera). The median detection limits are
given for different target spectral types (BAF, GK and M stars) and for
the $0.7\,''$ (\textit{solid line}) and $1.4\,''$ (\textit{dash dotted
line}) coronagraphic masks. \textit{Right:} VLT/NACO coronagraphic
detection limits in $K_s$-band (combined with the S27 camera). The
median detection limits are also given for different target spectral
types and coronagraphic masks.}

\label{fig:perfs_limdet}
\end{figure*}

In coronagraphy, the relative astrometry of the CCs
was obtained using a 2D-gaussian PSF fitting. The deconvolution
algorithm of V\'eran \& Rigaut (1998) and the maximization of the
cross-correlation function were applied using the primary star
(directly imaged) as PSF reference. The shifts ($\leq 1$ pixel)
induced between direct and coronagraphic images taken with different
filters, including neutral density, have been accounted for.  For the relative
photometry, classical aperture ($R_{ap} = 2 \times FWHM$) photometry
with residual sky-subtraction and classical deconvolution were used.
For faint sources detected at less than $\sim10~\sigma$, the
background subtraction become more critical and is responsible for
larger uncertainties in the deconvolution analysis. Our analysis was
then limited to a 2D-gaussian fitting coupled to aperture photometry
to derive the relative astrometry and photometry.

For observations obtained at several epochs, the proper motion and
parallactic motion of the primary star were taken into account to
investigate the nature of detected faint CCs. 
The relative positions recorded at different
epochs can be compared to the expected evolution of the 
position measured at the first epoch under the assumption that the
CC is either a stationary background object or a
comoving companion (see below). For the range of semi-major axes explored, any
orbital motion can be considered of lower order compared with the
primary proper and parallactic motions.

%________________________________
\subsection{Detection limits}

%
% Subsection Outlines
%
% - Detection limit estimation
%

The coronagraphic detection limits were obtained using combined direct
and coronagraphic images. On the final coronagraphic image, the
pixel-to-pixel noise was estimated within a box of 5$\times$5 pixels
sliding from the star to the limit of the NACO field of view. Angular
directions free of any spike or coronagraphic support contamination
were selected. Additionally, the noise estimation was calculated
within rings of increasing radii, a method which is more pessimistic
at close angular separation due to the presence of coronagraphic PSF
non-axisymmetric residuals. Final detection limits at $6\sigma$ were
obtained after division by the primary star maximum flux and
multiplication by a factor taking into account the ratio between the
direct imaging and coronagraphic integration times and the difference
of filter transmissions and bandwidths. Spectral type
  correction due to the use of different filters has been simulated
  and is smaller than 0.04~mag. The variation of the image quality
  (\textit{strehl ratio}) over the observation remains within 10\% and
  should not impact our contrast estimation from more than
  0.1~mag. The median detection limits, using the sliding box method,
  are reported in Fig.~\ref{fig:perfs_limdet}. They are given for
  observations obtained in $H$- and $K_s$-bands, with the
  $\oslash=0.7~\!''$ and $\oslash=1.4~\!''$ coronagraphic masks and
  for different target spectral types (BAF, GK and M stars) and will
  be used for the following statistical analysis of the survey.

\begin{figure}[t]
\centering
\includegraphics[width=\columnwidth]{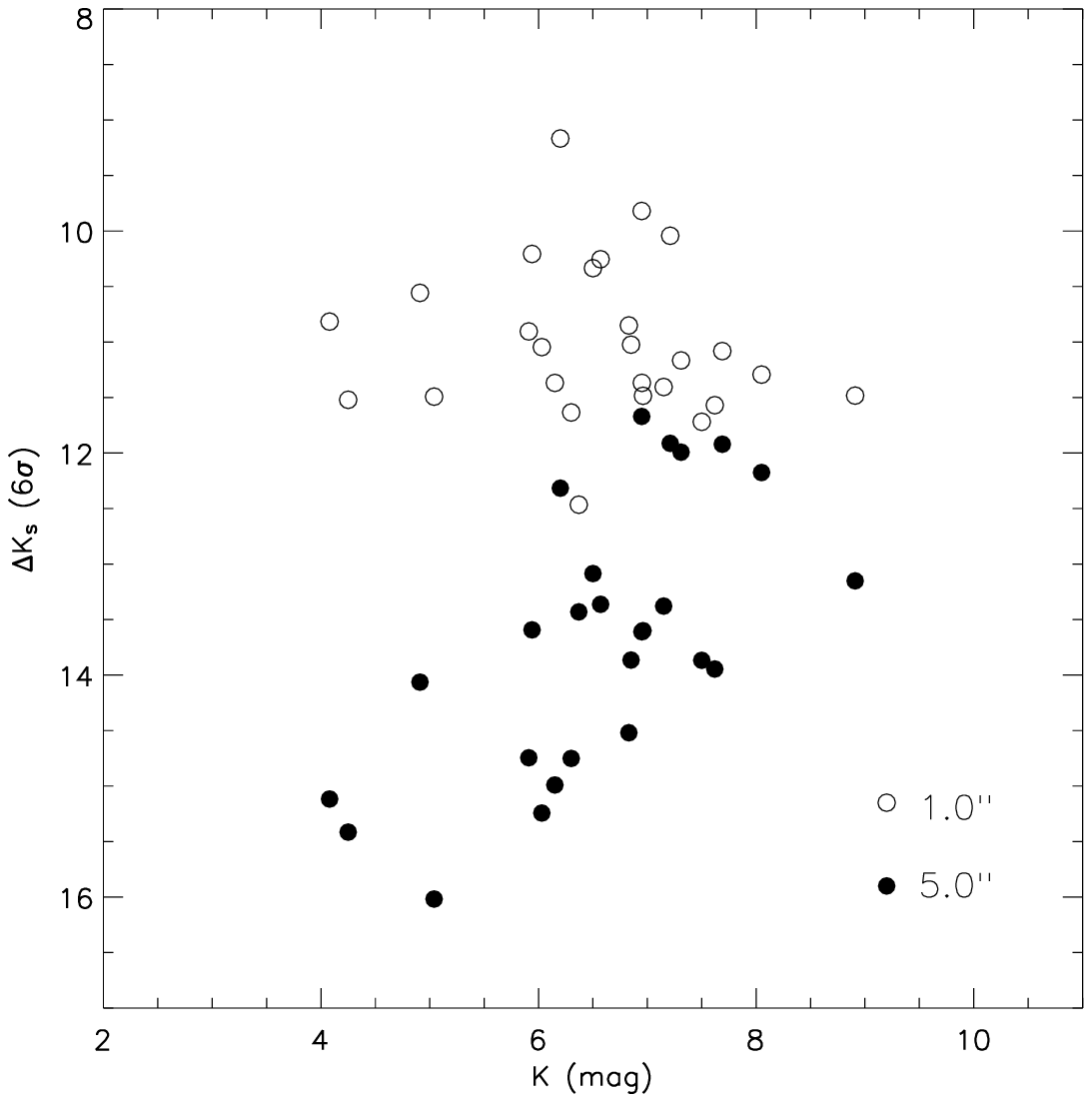}

\caption{VLT/NACO coronagraphic detection limits in $K_s$-band as a
function of the primary star brightness for two angular separations
(1.0~$\!''$ and 5.0~$\!''$). Two regimes can be seen; one at large
separations (shown here at $5.0\,''$) when the detection is limited by
detector read-out noise or background noise. The contrast varies then
linearly with the primary $K_s$ apparent magnitude due to the flux
normalization; a second regime at shorter separations (shown here at
$1.0\,''$) when the detection is speckle noise limited. Instrumental
quasi-static speckles are expected to dominate random, short-lived
atmospheric speckles and the contrast remains relatively constant over
a wide range of primary $K_s$ apparent magnitudes.}

\label{fig:perfs_limdet2}
\end{figure}

At large separations ($\ge 1.0-2.0\,''$) from the star when limited by
detector read-out noise or background noise, the contrast variation
with the primary spectral type is actually related to the primary nIR
brightness. This is shown in Fig.~\ref{fig:perfs_limdet2} in the case
of $K_s$-band detection limits at 5.0~$\!''$ as a function of the
primary $K_s$ apparent magnitude. The contrast varies linearly due to
the flux normalization. At shorter separations, the situation is more
complex as AO deep images are actually limited by the speckle noise.
Our detection limits remain constant over a wide range of
primary $K_s$ apparent magnitudes.
\begin{figure*}[t]
\centering
\includegraphics[width=\columnwidth]{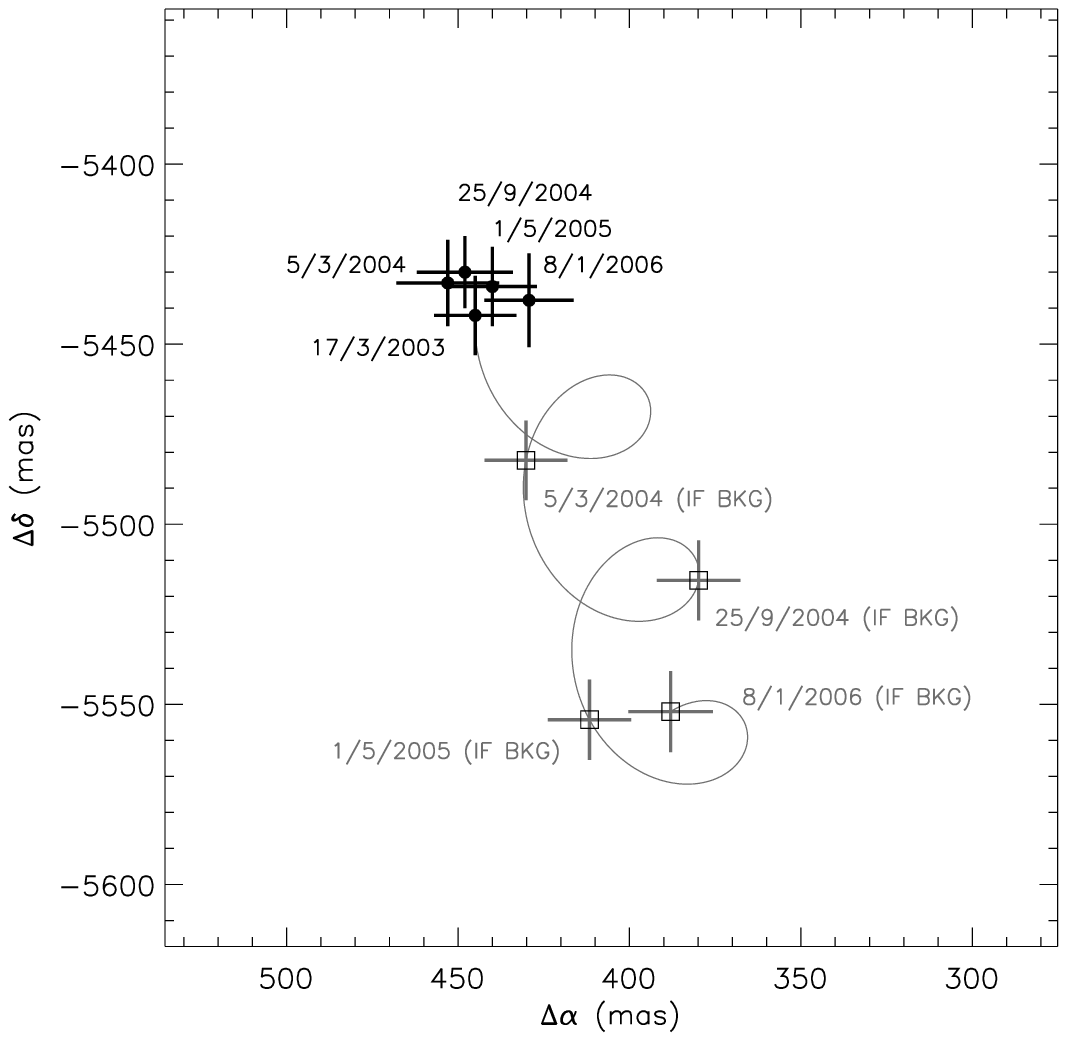}
\includegraphics[width=\columnwidth]{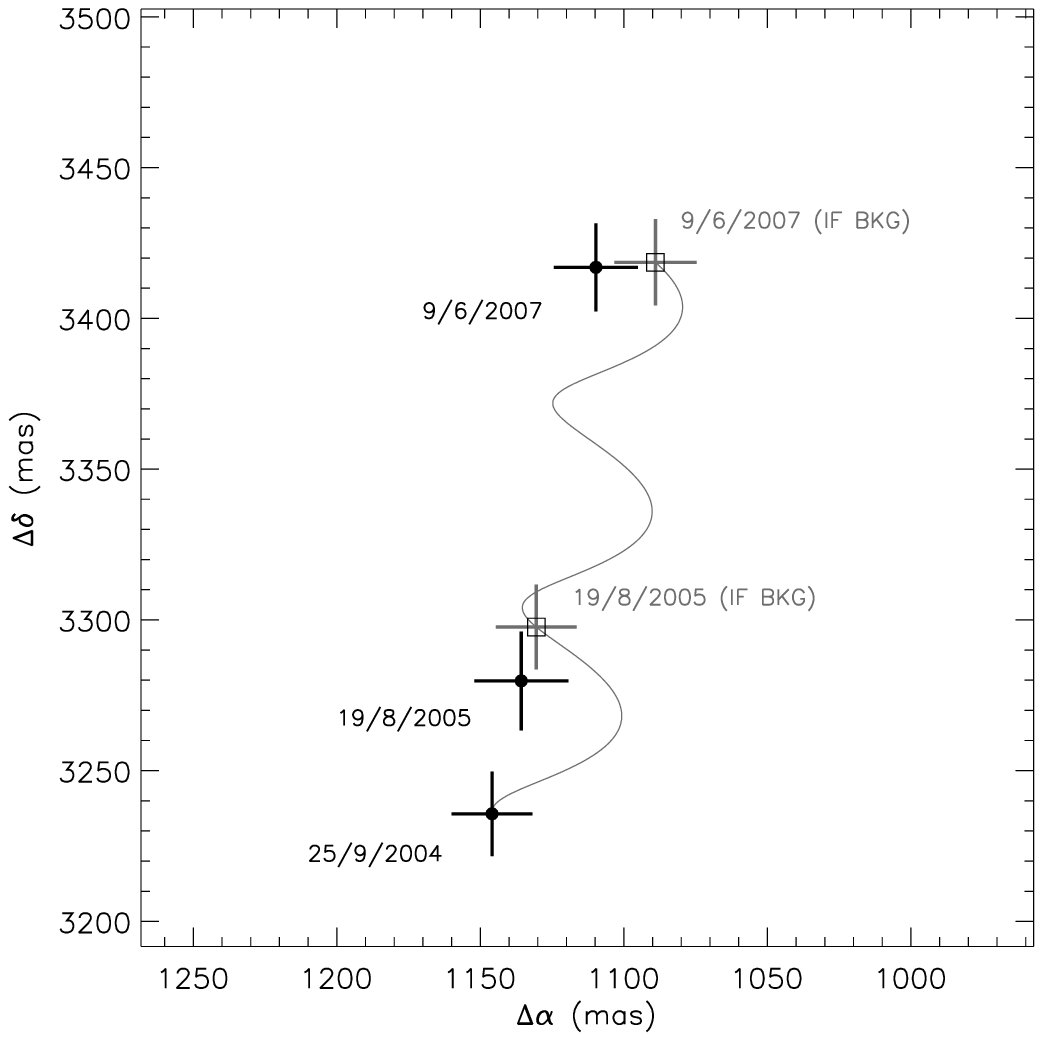}

\caption{VLT/NACO Measurements (\textit{full circles} with
uncertainties) of the offset positions of the comoving companion AB
Pic\,b to A (\textit{left}) and of the CC relative to 0ES1847
(\textit{right}). For each diagram, the expected variation of
offset positions, if the candidate is a background object, is shown
(\textit{solid line}). The variration is estimated based on the
parallactic and proper motions of the primary star, as well as the
initial offset position of the CC from A. The \textit{empty boxes}
give the corresponding expected offset positions of a background
object for the different epochs of observations (with
uncertainties). In the case of AB\,Pic\,b, the relative positions do
not change with time confirming that AB\,Pic b is comoving. On the
contrary, the relative position of the CC to 0ES1847 varies in time
as predicted for a background stationary contaminant. For our sample,
astrometric follow-up over 1-2 years enabled a rapid identification.}

\label{fig:followup}
\end{figure*}

All published deep imaging surveys dedicated to planet search
(Masciadri et al. 2005; Kasper et al. 2007; Lafreni\`ere et al. 2007;
Biller et al. 2007), including this one, derived detection threshold
assuming that the residual noise in the final processed image follows
a Gaussian intensity distribution. A typical detection threshold at 5
or 6~$\sigma$ is then usually assumed over the complete angular
range. Whereas the approximation of a Gaussian distribution for the
residual noise is valid within the detector read-out noise or
background noise regime, careful analysis by Marois et al. (2008a)
shows that this is not adequate at small separations when speckle
noise limited (typically $\le 1.0-2.0\,''$ in our survey; see
Fig.~4). In this regime, AO deep images are actually not limited by
random, short-lived atmospheric speckles, but by instrumental
quasi-static speckles.  A non-gaussian distribution of the residual
noise must be taken into account to specify a detection threshold at a
given confidence level.  Therefore, our current 6~$\sigma$ detection
threshold at small separations is probably too optimistic.  However,
the systematic error induced in our sensitivity limits is probably of
lower significance than uncertainties in planet age and use of
uncalibrated planet evolutionary models as described below.

%
%__________________________________________________________________
\section{Results}

%
% Data Reduction and Analaysis
%
% - Status of new multiple systems
%   1. New visual ones
%   2. Confirmed comoving ones
%   3. Tight calibrator TWA22 
% - Status of companion candidates
% - Detection performances
%

The main purpose of our survey was the detection of close brown dwarf
and planetary mass companions using the deep imaging technique on an
optimized sample of nearby stars. Compared to other works, our
strategy has been sucessful with the confirmation of the brown dwarf
companion to GSC\,08047-00232 (Chauvin et al. 2003; 2005a) and the
discoveries of one planetary mass companion, around the young
brown dwarf 2MASSW J1207334-393254 (hereafter 2M1207; Chauvin et
al. 2004; 2005c) and one companion at the planet/brown dwarf
  boundary to the young star AB Pic (Chauvin et al. 2005b).

In this section, we detail the three main results of this survey:
\begin{enumerate}
\item the identification of a large fraction of contaminants in the
close angular environment of our sample of young, nearby stars. This
identification step is necessary for the statistical analysis of our
complete set of detection limits (see below). Contaminants
identification serves in addition to the preparation of future
deep imaging search of exoplanets that will re-observe most of these
stars.
\item the discovery of several new close stellar multiple systems,
despite our binary rejection process. Three systems are actually
confirmed to be comoving. One is a possible low-mass calibrator for
the predictions of evolutionary models.
\item Finally, we review the status of the three substellar
companions, confirmed with NACO, in regards of the latest results in
the literature and from our survey.
\end{enumerate}

%________________________________
\subsection{Contaminant identification}

%
% Subsection Outlines
%
% - How many faint sources detected
%   . total of systems with companion candidates
%   . number of comapnion candidates detected
% - Number summary: 
%   . number of systems observed at 1 epoch with VLT.
%   . number of systems observed at several eopchs with VLT.
%   . number of systems observed at several eopchs with VLT and HST.
% - Relative position and photometry results
% - Status (significance level)
% - Example of confirmed companion and false alarm with Twa19~A. 
%

Among the complete sample composed of 88 stars, a total of 65 were
observed with coronagraphic imaging. The remaining 23 targets were
observed in direct or saturated imaging because the system was
resolved as a $1.0-12~\!''$ visual binary inappropriate for deep
coronagraphic imaging, because the atmospheric conditions were
unstable or because the system was simply too faint to warrant
efficient use of the coronagraphic mode.

Among the 65 stars observed with both direct imaging and
coronagraphy, nothing was found around 29 (45\%) stars and at least
one CC was detected around the 36 (55\%) others. A total of $\sim236$
CCs were detected. To identify their nature, 14 (39\%) systems were
observed at two epochs (at least) with VLT and 16 (44\%) have combined
VLT and HST observations at more than a one year interval (Song et
al. 2009, in prep). Finally, 6 (17\%) were observed at only one epoch
and require further follow-up observations. The position and
photometry of each detected CC relative to its primary star, at each
epoch, are given in Tables 7-13. For multi-epoch observations, to
statistically test the probability that the CCs are background objects
or comoving companions, a $\chi^2$ probability test of $2\times
N_{epochs}$ degrees of freedom (corresponding to the measurements:
separations in the $\Delta\alpha$ and $\Delta\delta$ directions for
the number $N_{epochs}$ of epochs) was applied. This test takes into
account the uncertainties in the relative positions measured at each
epoch and the uncertainty in the primary proper motion and parallax
(or distance). Fig.~\ref{fig:followup} gives an illustration of a
($\Delta\alpha$, $\Delta\delta$) diagram that was used to identify a
stationary background contaminant near 0ES1847. A status of each CC
has been assigned as confirmed companion (C; P$_{\chi^2}<0.1$\%), background
contaminant (B; P$_{\chi^2}>99$\%), probably background (PB;
P$_{\chi^2}>99$\%, but combining data from two different instruments)
and undefined (U). Over the complete coronagraphic sample, 1\% of the
CCs detected have been confirmed as comoving companions, 43\% have
been identified as probable background contaminants and about 56\%
need further follow-up observations. The remaining CCs come mostly
from the presence of background crowded fields in the field of view of
the 6 stars observed at one epoch.

Among the 23 stars and brown dwarfs observed only in direct or
saturated imaging, several have been resolved as tight multiple
systems (see below). 4 stars (FS1174, FS979, FS1017 and FS1035) have
at least one substellar CC (see
Tables~\ref{tab:faint_candidates4},~\ref{tab:faint_candidates5}
and~\ref{tab:faint_candidates6}). FS1035 was observed at two
successive epochs and the faint object detected at $\sim5.6~\!''$ has
been identified as a contaminant.

\begin{figure}[t]
\includegraphics[width=\columnwidth]{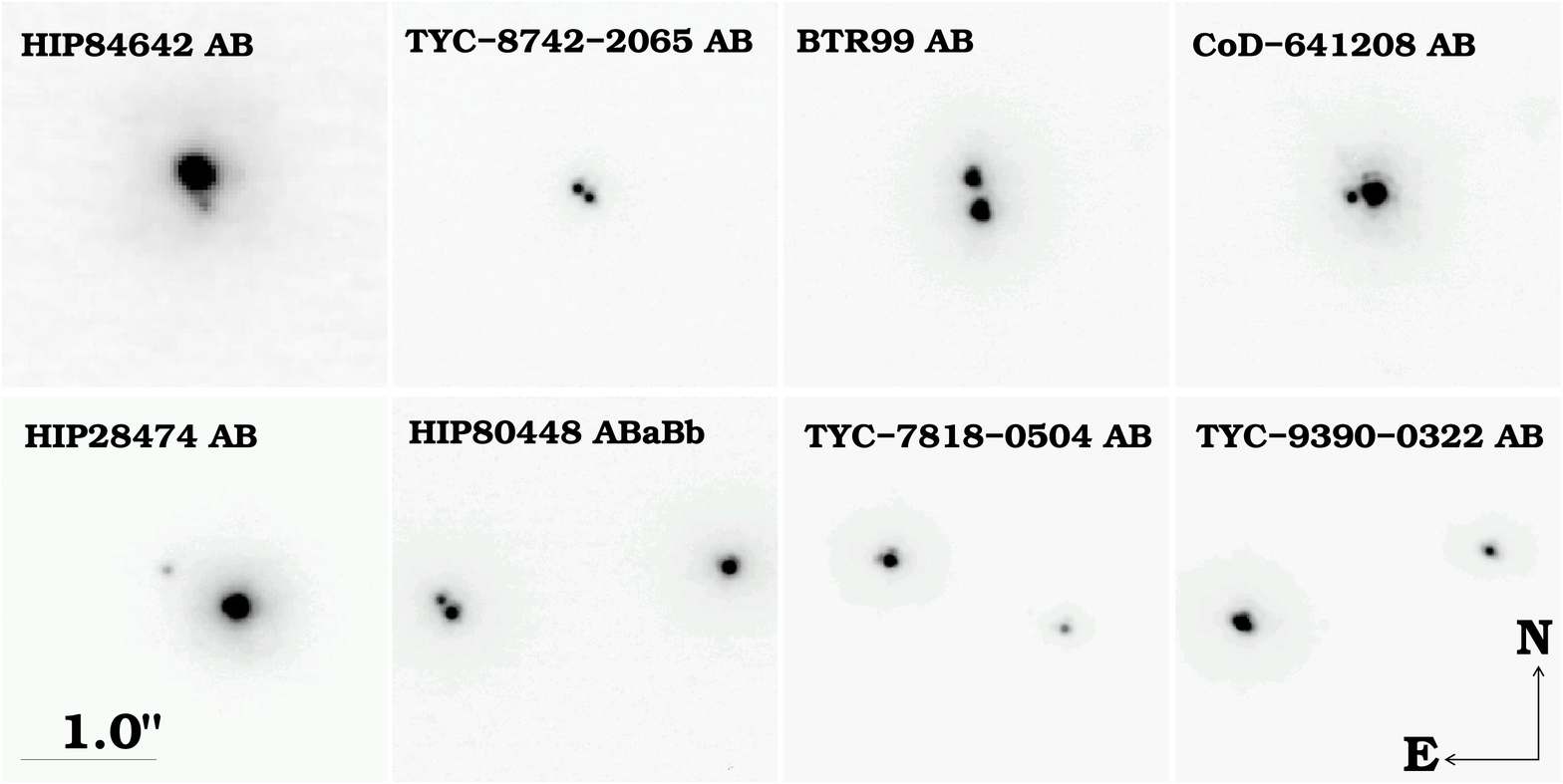}
\vspace{0.1cm}\\
\includegraphics[width=\columnwidth]{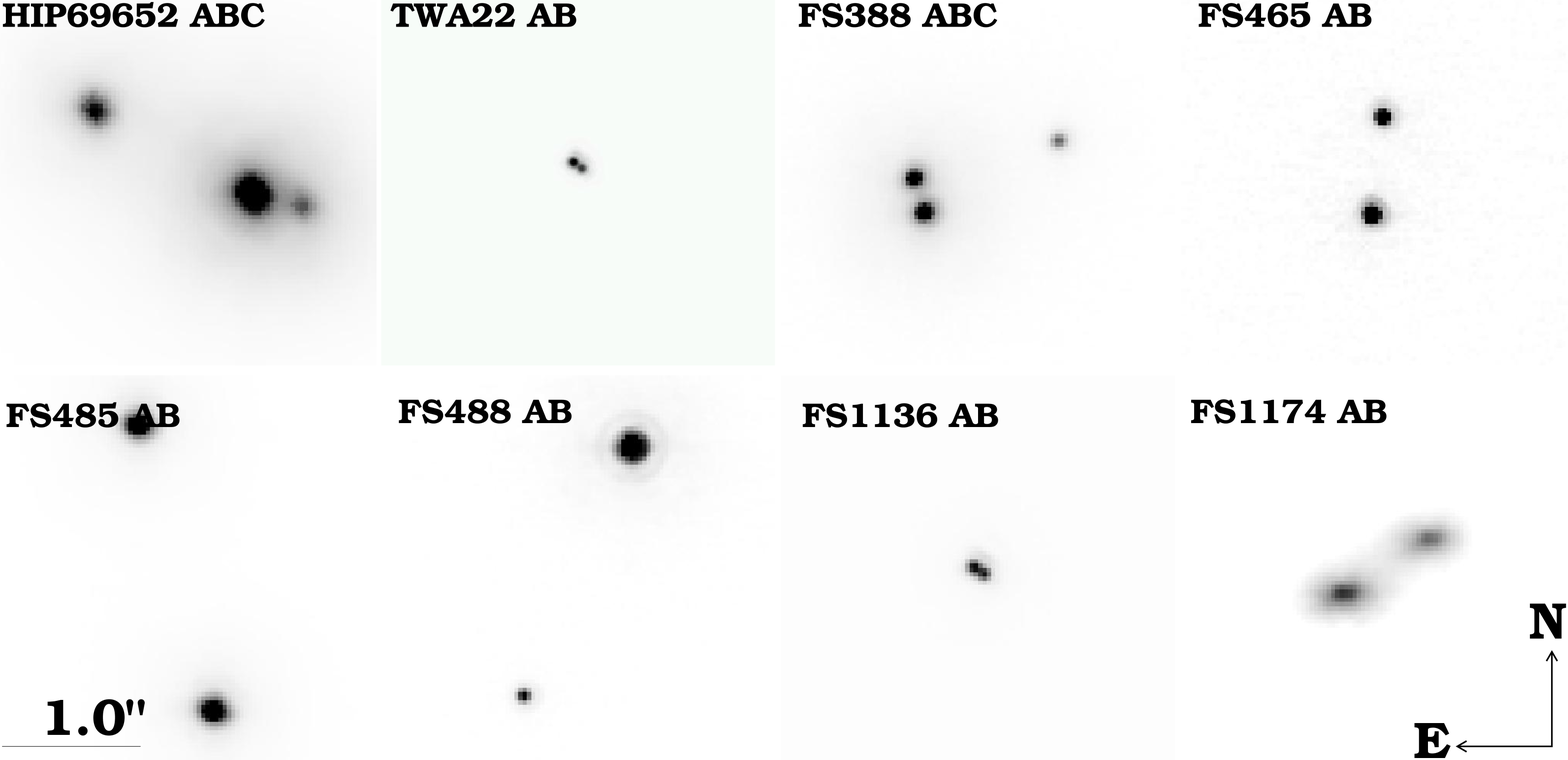}
\vspace{0.1cm}\\
\caption{New visual binaries resolved with NACO at
VLT. HIP\,108195\,AB, HIP\,84642\,AB and TWA22\,AB were in addition
confirmed as comoving multiple systems. TWA22\,AB was 
monitored for 4 years to constrain the binary orbit and determine its total
dynamical mass (see Bonnefoy et al. 2009, accepted)}

\label{fig:new_binaries}
\end{figure}

%________________________________
\subsection{Close stellar multiple systems}

\subsubsection{New visual binaries}

Our survey was not aimed at detecting new stellar binaries. Known
bright equal-mass binaries of $1.0-12.0~\!''$ separation were rejected
from our sample as they degrade the coronagraphic detection
performances by limiting dynamical range. A few tight binaries were
kept when both components could be placed behind the coronagraphic
masks. Despite our binary rejection process, 17 new close visual
multiple systems were resolved (see Fig.~\ref{fig:new_binaries} and
\ref{fig:fig1081095}). They include 13 tight resolved binaries and 4
triple systems.  Their relative flux and position are reported in
Table~\ref{tab:new_binaries}. Their separations range between
$0.1-5.0~\!''$ and their $H$ and $K_s$ contrasts between
$0.0-4.8$~mag. Among them, HIP\,108195\,ABC, HIP\,84642\,AB and
TWA22\,AB were observed at different epochs and are confirmed as
comoving systems.

\subsubsection{The comoving multiple systems HIP\,108195\,ABC and HIP\,84642\,AB}

\begin{figure}[t]
\centering
\includegraphics[width=7cm]{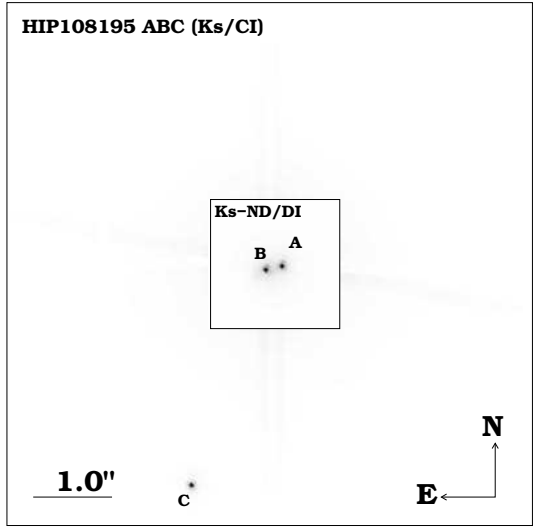}
\caption{Composite VLT/NACO Ks-band image of the triple comoving
system HIP108195\,ABC. The inner part shows the direct image (DI)
of HIP108195\,AB (attenuated by a factor $\sim100$) using the $K_s$-band with a neutral density filter to
avoid saturation. Both components of the astrometric binary cataloged
by Hipparcos are resolved. The outer part shows the deeper
coronagraphic image obtained in $K_s$-band with the C component about 100 times fainter than A or B.}
\label{fig:fig1081095}
\end{figure}

Close to the Hipparcos double star HIP\,108195\,AB (F3, 46.5~pc),
member of Tuc-Hor, we resolved a faint source at $4.96\,''$
($\Delta_{proj}=230$~AU; i.e $a\sim290$~AU). In addition to a
confirmation that HIP\,108195\,AB is a comoving pair, we found that
the fainter source is a third component of this comoving multiple
system (Fig.~\ref{fig:fig1081095}). Combined distance, age and
apparent photometry are compatible with an M5-M7 dwarf according to
PMS model predictions (Siess et al. 2000) and places the companion at
the stellar/brown dwarf boundary.

HIP\,84642 (K0, 58.9~pc) is not reported as a double star in the
Hipparcos Visual Double Stars catalog (Dommanget et al. 2000),
possibly due to the small angular separation and the flux ratio with
the visual companion.  Based on combined VLT/NACO images from our
programme and from the SACY survey (Hu\'elamo et al. 2009, in prep),
we confirm that the companion shares common proper motion with
HIP\,84642. This object is likely to be an M4-M6 young dwarf comparing
its photometry to predictions of PMS models. Based on the statistical
relation between projected separation and semi-major axis of Couteau
(1960), HIP\,84642\,AB is likely to be a tight ($\Delta_{proj}=14$~AU;
$a\sim18$~AU; K0-M5) binary with a period of several tens of years.

\begin{table}[t]
\caption{Relative positions and $K_s$ and $H$-band contrast of the new
binaries resolved by NACO at VLT. Contrast uncertainty is about 0.1~mag.}
\label{tab:new_binaries}      
\begin{tabular*}{\columnwidth}{@{\excs}llllll}     % 7 columns 
\hline\hline\noalign{\smallskip}       
Name                    & UT Date      &  $\Delta$       & P.A.              & $\Delta K_s$              \\ 
                        &              &  (mas)       & (deg)                  & (mag)                     \\
\noalign{\smallskip}\hline\noalign{\smallskip}
HIP108195\,AB           & 19-08-2005   & $339\pm5$    & $102.7\pm0.2$    & $0.0$                 \\
\hspace{1.43cm} AC      & 19-08-2005   & $4964\pm9$   & $158.4\pm0.2$    & $4.8$                 \\
HIP84642\,AB            & 05-06-2007   & $220\pm14$   & $191.3\pm0.8$    & $2.5$                 \\
TYC-8742-2065\,AB       & 27-04-2004   & $114\pm2$    & $232.5\pm0.4$    & $0.2$                 \\
BTR99\,AB               & 25-09-2004   & $264\pm3$    & $12.8\pm0.3$     & $0.5$                 \\
CoD-641208\,AB          & 25-09-2004   & $178\pm3$    & $95.3\pm0.4$     & $2.3$                 \\
HIP80448\,BaBb          & 27-04-2004   & $134\pm2$    & $37.5\pm0.7$     & $1.1$                 \\            
HIP69652\,AB            & 26-02-2006   & $319\pm6$    & $256.8\pm0.5$    & $1.3$                 \\
\hspace{1.28cm} AC      & 26-02-2006   & $1123\pm6$   & $61.8\pm0.4$     & $0.8$                 \\
TWA22\,AB               & 05-03-2004   & $100\pm5$    & $80.2\pm0.2$     & $0.4$                 \\
FS388           AB      & 08-01-2006   & $224\pm5$    & $16.4\pm0.6$     & $0.3$                 \\            
\hspace{0.85cm} AC      & 08-01-2006   & $963\pm6$    & $297.7\pm0.2$    & $1.6$                 \\            
FS465\,AB               & 08-01-2006   & $619\pm6$    & $353.4\pm0.3$    & $0.7$                 \\
FS488\,AB               & 08-01-2006   & $1710\pm7$   & $156.6\pm0.1$    & $2.8$                 \\
FS485\,AB               & 26-02-2006   & $1862\pm7$   & $14.7\pm0.1$     & $0.2$                 \\
FS1136\,AB              & 19-08-2005   & $74\pm5$     & $122.0\pm0.8$    & $0.2$                 \\
FS1174\,AB              & 19-08-2005   & $626\pm6$    & $302.7\pm0.5$    & $0.4$                 \\
%2M1207\,AB              & 19-08-2005   & $772\pm4$    & $125.4\pm0.3$    & $0.2$                 \\
%2M0652                  & 08-01-2006   & $227\pm6$    & $306.7\pm0.8$    & $0.4$                 \\
\noalign{\smallskip}\hline\noalign{\smallskip}
\noalign{\smallskip}\hline                  
\end{tabular*}

\begin{tabular*}{\columnwidth}{@{\excs}llllll}     % 7 columns 
Name                    & UT Date      &  $\Delta$       & P.A.              & $\Delta H$              \\ 
                        &              &  (mas)       & (o)                  & (mag)                     \\
\noalign{\smallskip}\hline\noalign{\smallskip}
HIP28474\,AB            & 27-04-2004   & $613\pm3$    & $61.7\pm0.2$     & $3.8$                 \\
TYC-7818-0504\,AB       & 17-03-2003   & $1463\pm6$   & $111.3\pm0.2$    & $1.7$                 \\
TYC-9390-0322\,AB       & 17-03-2003   & $2005\pm7$   & $286.3\pm0.1$    & $1.6$                 \\
%Recherche Catalogue
%Detected in 2MASS or in Hipparcos
%FS979                   & 19-08-2005   & $7464\pm21$  & $119.2\pm0.1$    & $0.5$                 \\
%HIP76768                & 19-08-2005   & $825\pm6$    & $49.0\pm0.3$     & $0.6$                 \\
%HIP107705               & 27-04-2004   & $1208\pm4$   & $41.4\pm0.2$     & $0.6$                 \\
%CoD-33386               & 22-11-2002   & $2981\pm6$   & $111.1\pm0.1$    & $0.0$                 \\
\noalign{\smallskip}\hline                  
\end{tabular*}
\end{table}

\subsubsection{The young, tight astrometric binary TWA22\,AB}

The tight ($\sim$100~mas; $a\sim1.8$~AU) binary TWA22\,AB was observed at several
epochs. We aimed at monitoring the system orbit to determine the total
dynamical mass of this system using an accurate distance determination
($17.53\pm0.21$~pc, Texeira et al. 2009, submitted). The physical properties
(luminosity, effective temperature and surface gravity) of each
component were obtained based on near-infrared photometric and
spectroscopic observations. By comparing these parameters with
evolutionary model predictions, we consider the age and the
association membership of the binary. A possible under-estimation of
the mass predicted by evolutionary model for young stars close to the
substellar boundary is presented in two dedicated papers (Bonnefoy
et al. 2009, accepted; Texeira et al. 2009, accepted).

%________________________________
\subsection{Substellar companions}

We review below the latest results about the three substellar
companions GSC\,08047-00232\,B, AB\,Pic b and 2M1207\,b since their
initial companionship confirmation. Recent age, distance, astrometric and
spectroscopic measurements enable us to refine their predicted physical
properties and their origin in regards to other confirmed substellar
companions in young, nearby associations.

\subsubsection{GSC\,08047-00232\,B}

Based on the ADONIS/SHARPII observations of two dozen probable
association members of Tuc-Hor, Chauvin et al. (2003) identified a
$20\pm5$~\Mjup\ candidate to GSC\,08047-00232 (CoD-52381). This
candidate was independently detected by Neuh\"auser et al. (2003) with
the SHARP instrument at the ESO \textit{New Technology Telescope}
(NTT). Neuh\"auser \& Guenther (2004) acquired $H$- and $K$-band
spectra and derived a spectral type M$8\pm2$, corroborated by Chauvin
et al. (2005a). Finally, in the course of our VLT/NACO observations,
we confirmed that GSC\,08047-00232\,B was comoving with A (Chauvin et
al. 2005a).  Mass, effective temperature and luminosity were
determined by comparing its $JHK$ photometry with evolutionary model
predictions and the Tuc-Hor age and photometric distance for the
system.  The results are reported in Table~7 and compared to the
complete list of confirmed substellar companions discovered among the young,
nearby associations.  Membership in Tuc-Hor and the assigned age of
GSC\,08047-00232\,AB have been debated for a time. Further studies of
loose young associations sharing common kinematical and physical
properties recently led Torres et al. (2008) to identify
GSC\,08047-00232\,AB as a high-probability ($80\%$) member of the
Columba association of age 30~Myr, confirming the young age and the
brown dwarf status of GSC\,08047-00232\,B.
 
\subsubsection{AB Pic\,b}

During our survey, a $13\pm2$~\Mjup\ companion was discovered near the
young star AB\,Pic (Chauvin et al. 2005b). Initially identified by
Song et al. (2003) as a member of Tuc-Hor, the membership of AB Pic
has been recently discussed by Torres et al. (2008) who attached this
star to the young ($\sim$30~Myr) Columba association. Additional
astrometric measurements of the relative position of AB Pic\,b to A
firmly confirm the companionship reported by Chauvin et al. (2005b;
see Fig.~\ref{fig:followup}, \textit{left panel}). Based on age, distance
and nIR photometry, Chauvin et al. (2005b) derived the physical
properties of AB Pic\,b based on evolutionary models (see Table~7). As
per the three young substellar companions to TWA5A, HR7329 and
GSC\,08047-00232, AB Pic\,b is located at a projected physical
separation larger than 80~AU. Formation by core accretion of
planetesimals seems unlikely because of inappropriate timescales to
form planetesimals at such large distances. Gravitational
instabilities within a protoplanetary disk (Papaloizou \& Terquem
2001; Rafikov 2005; Boley 2009) or Jeans-mass fragmentation proposed for brown
dwarf and stellar formation appear to be more probable pathways to
explain the origin of the Table~7 secondaries.

\begin{table*}[t]
\label{tab:subcomp}
\caption{Properties of the confirmed comoving substellar companions discovered
in the young, nearby associations: TW Hydrae (Twa), $\beta$ Pictoris
($\beta$ Pic), Columba (Col) and Carina (Car). Tentative spectral type have
been determined from nIR spectroscopic observations, whereas masses
and effective temperatures are predicted by evolutionary models
based on the nIR photometry, the age and the distance of the system. ($^*$), for 2M1207\,b, Mohanty et al. (2007) suggests a higher mass of
$8\pm2$~M$_{\rm{Jup}}$ and the existence of a
circum-secondary edge-on disk to explain their measured effective
spectroscopic temperature of $1600\pm100$~K.}
\begin{tabular*}{\textwidth}{@{\excs}llllllllll}
\hline\hline\noalign{\smallskip}
Name              &      Grp            &  Age       &   $d$                     &  SpT$_A$       &  SpT$_B$       &   M$_B$                  &   $T^B_{\rm{eff}}$  &  $q_{B/A}$  & $\Delta_{proj}$ \\
                  &                     &  (Myr)     &   (pc)                  &                &                &  ($\rm{M}_{\rm{Jup}}$)   & (K)                 &         &  (AU)        \\ 
\noalign{\smallskip}\hline\noalign{\smallskip}
TWA5              &  TWA                &  8         & (45-50)                    
& M1.5           & M8.5           & $25\pm5$                 &    $2500\pm150$     & 0.055   & $93\pm10$       \\
HR7329            &  $\beta$ Pic        &  12        & 48.2$^{+1.8}_{-1.6}$    & A0V            & M8             & $25\pm5$                 &    $2550\pm150$     & 0.010   & $199\pm10$    \\ 
\noalign{\smallskip}
GSC-08047-00232          &  Tuc-Hor/Col    &  30        & (85-95)                    & K3V            & M$9^{+1}_{-3}$ & $20\pm5$                 &    $2100\pm200$     & 0.025   & $295\pm30$    \\
AB Pic            &  Tuc-Hor/Car        &  30        & 46.0$^{+1.6}_{-1.5}$    & K2V            & L1$^{+2}_{-1}$ & $13\pm2$                 &    $1700\pm200$     & 0.015   & $250\pm10$    \\
2M1207            &  TWA                &  8         & 52.4$^{+1.1}_{-1.1}$    
& M8             & late-L         & $4\pm1^*$                  &    $1150\pm150^*$     & $0.16^*$    & $40\pm2$      \\   
%                  &                     &            &                         &                &                & $8\pm2$                  &    $1600\pm100$     & 0.32    &                \\   
\noalign{\smallskip}\hline
\end{tabular*}
\end{table*}

\subsubsection{2M1207\,b}

Among the young candidates of our sample, a small number of very low
mass stars and brown dwarfs were selected to take advantage of the
unique capability offered by NACO at VLT to sense the wavefront in the
IR. Most were observed in direct and saturated imaging. This strategy
proved to be successful with the discovery of a planetary mass
companion in orbit around the young brown dwarf 2M1207 (Chauvin et
al. 2004; 2005c). HST/NICMOS observations independently confirmed this
result (Song et al. 2006). A low signal-to-noise spectrum in H-band
enabled Chauvin et al. (2004) to suggest a mid to late-L dwarf
spectral type, supported by its very red nIR colors. Additional low
signal-to-noise spectroscopic observations compared with synthetic
atmosphere spectra led Mohanty et al.  (2007) to suggest an effective
spectroscopic temperature of $1600\pm100$~K and a higher mass of
$8\pm2$~M$_{\rm{Jup}}$. To explain the companion under-luminosity,
Mohanty et al. (2007) have suggested the existence of a
circum-secondary edge-on disk responsible for a gray extinction of
$\sim2.5$~mag between 0.9 and 3.8~$\mu$m. However, synthetic
atmosphere models clearly encounter difficulties in describing
faithfully the late-L to mid-T dwarfs transition ($\sim1400$~K for
field L/T dwarfs), corresponding to the process of cloud
clearing. Similar difficulties have been encountered by Marois
  et al. (2008b) to reproduce all photometric data of the three
  planetary mass companions to HR\,8799 that fall also at the edge or
  inside the transition of cloudy to cloud-free atmospheres. In the
case of 2M1207\,b, future spectroscopic or polarimetric observations
should help to distinguish between the two scenarios (obscured or
non-obscured by a circumstellar disk).  Recent precise parallax
determinations (Gizis et al. 2007; Ducourant et al. 2008) allowed a
reevaluation of the distance and the physical properties of the
companion (see Table~7).

%
%__________________________________________________________________
\section{Statistical analysis}

% Plan
% 1. Introduction, different surveys. Description.
% - Conversion in terms of mass (Fig,; Histograms?)
% - Two possible approaches to answer to different questions
%   . RV as input distributions. Extrapolation 
%   . Planet Fraction Upper Limit
%
% Figures (1): LiMassePlus, Histogramme
%
% 2. Input Ditributions
%
% Figures (1, 2?): Variation with different parameters:
% period, mass, primary mass dependency 
%
% 3. Planet fraction
%
% Figures (2): Mean detection Probability (uniform distributions)
% Planet Fraction Upper Limit
%
% 4. Analysis limitations:
%  . evolutionary models?
%  . extrapolation of input distribution? evolution
%  . planetray mass scaling?
%  . Contrast estimation? Marois et al.
%

%________________________________
\subsection{Context}

% 1. Introduction, different surveys. Description.
% - Conversion in terms of mass (Fig,; Histograms?)
% - Two possible approaches to answer to different questions
%   . RV as input distributions. Extrapolation 
%   . Planet Fraction Upper Limit
%

Over the past years, a significant number of deep imaging surveys have
been reported in the literature, dedicated to the search for
exoplanets around young, nearby stars (Chauvin et al. 2003;
Neuh\"auser et al. 2003; Lowrance et al. 2005; Masciadri et al. 2005;
Biller et al. 2007; Kasper et al. 2007; Lafreni\'ere et
al. 2007). Various instruments and telescopes were used with different
imaging techniques (coronagraphy, angular or spectral differential
imaging, $L'$-band imaging) and observing strategies. None of those
published surveys have reported the detection of planetary mass
companions that could have formed by a core-accretion model (as
expected for a large fraction of planet candidates reported by RV
measurements).  Several potential planetary mass companions were
discovered, but generally at relatively large physical separations or
with a small mass-ratio with their primaries, suggesting a formation
mechanisms similar to (sub)stellar binaries and stars. Only very
recently, planet candidates perhaps formed by core-accretion have been
imaged around the stars Fomalhaut (Kalas et al. 2008), HR\,8799
(Marois et al. 2008b) and $\beta$ Pictoris (Lagrange et al. 2009),
initiating the study of giant exo-planets at the (mass, distance)
scale of our solar system.

Confronted with a null-detection of planets formed by core-accretion,
several groups (Kasper et al. 2007; Lafreni\`ere et al. 2007; Nielsen et
al. 2008) have developed statistical analysis tools to exploit the
complete deep imaging performances of their surveys. A first approach
is to test the consistency of various sets of (mass, eccentricity,
semi-major axes) parametric distributions of a planet population in
the specific case of a null detection. A reasonable assumption is to
extrapolate and normalize planet mass, period and eccentricity
distributions using statistical results of RV studies at short
periods. Given the detection perfomances of a survey, the rate of
detected simulated planets (over the complete sample) enables
derivation of the probability of non-detection of a given planet
population associated with a normalized distribution set.  Then
comparison with a survey null-detection sample tests directly the
statistical significance of each distribution and provides a simple
approach for constraining the outer portions of exoplanetary systems.

A second more general approach aims at actually constraining the
exoplanet fraction $f$ within the physical separation and mass probed
by the survey, in the case of null or positive detections. Contrary to
what was assumed before, $f$ becomes an output of the simulation, that
actually depends on the assumed (mass, period, eccentricity)
distributions of the giant planet population.  This statistical
analysis aims at determining $f$ within a confidence interval as a
function of mass and semi-major axis, given a set of individual
detection probabilities $p_j$ directly linked to the detection limits
of each star observed during the survey and the considered giant
planet distributions. One can refer to the work of Lafreni\`ere et
al. (2007) and Carson et al. (2006), for a general description of the
statistical formalism applied for this analysis.

For our survey, we will consider the specific case of a null detection
of planet formation by core-accretion within a Poisson statistical
formalism that leads to a simple analytical solution for the exoplanet
fraction upper limit ($f_{\rm{max}}$). In the following, we will
consider both approaches to exploit the full survey detection
potential.

\begin{figure}[t]
\centering
\includegraphics[width=\columnwidth]{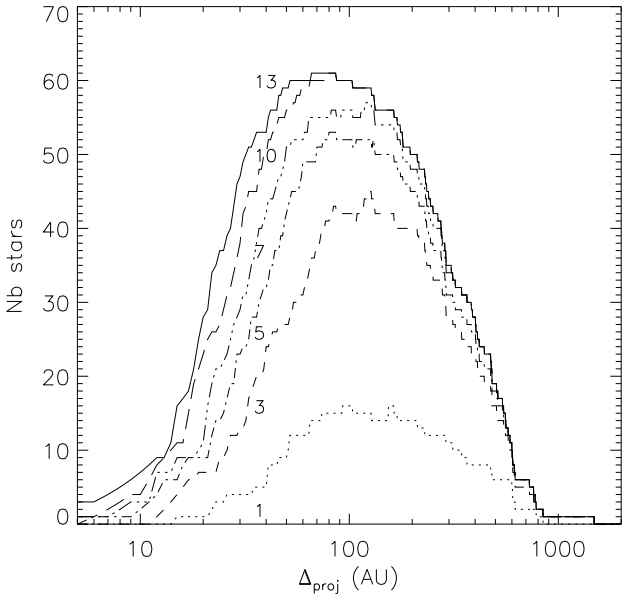}

\caption{Histogram of projected physical separations explored, for
various planetary masses (1, 3, 5, 7, 10 and 13)~M$_{\rm{Jup}}$, in
the close vicinity of the 65 young, nearby stars observed with NACO at
VLT in coronagraphy. Contrast performances have been converted into
masses based on the nIR photometry, age and distance of the primary
stars.}
\label{fig:masse_physdist}
\end{figure}

%________________________________
\subsection{Simulation description}

The simulation process is similar to the one adopted by Kasper et
al. (2007), Lafreni\`ere et al. (2007) and Nielsen et al. (2008). Due to
the important spectral type dispersion of our sample, we have included
in addition a planet mass dependency on primary mass. The
different steps of the simulation process are described below:

\begin{enumerate}

\item Our simulation star sample is composed of 65 stars observed in
coronagraphic imaging mode (see Table~2 and 3). Binaries that could
impact the presence of a planet within a range of semi-major axis of
$a = [5-150]$~AU were removed. Apparent magnitude, distance, age and
mass are the prime simulation parameters.

\item The detection limits were converted to predicted masses
using COND03 and DUSTY evolutionary models of Chabrier et al. (2000)
and Baraffe et al. (2003). COND03 models are adapted to predict
properties of cool ($\le1700$~K) substellar objects, whereas DUSTY
model predictions were considered for hotter temperatures. Based on
our (6$\sigma$) individual detection limits and target (distance, age,
$H$ or $K_s$-band magnitude) properties, we derived the space of
predicted masses and projected physical separation explored around
each star of the sample (see histogram in
Fig.~\ref{fig:masse_physdist}).

\item For the giant planet population, we have considered input
distributions based on parametric laws for mass and period
extrapolated from RV studies. The eccentricity distribution was chosen
to follow the empirical planet distributions of RV planets. For mass
and period, we consider power laws $dN/dM_psini \propto (M_psini)^\alpha$
and $dN/dP \propto P^\beta$ respectively. In addition, the influence
of a planetary mass distribution scaled as a function of the stellar
mass ($M_p \propto M_*^\gamma$) was tested.

\item Monte Carlo simulations were run to take into account the the
exoplanet distributions and orbital phase. For each run, 10,000 values
of $M.\rm{sin}i$ and $P$ are randomly generated, following the adopted
exoplanet distributions, together with all the other orbital elements,
which are supposed to be uniformly distributed.  The real
characteristics of each target star (mass, distance) are taken into
account to evaluate the semi-major axis and projected physical
separation of the planets.

\item The final step is a comparison with the survey null-detection results
and detection performances: either for a derivation of a non-detection
probability and thus constraining the statistical significance of
various sets of input distributions or for a derivation of the planet
fraction upper limit ($f_{\rm{max}}$) for a given set of exoplanet
distributions. Dead zones of our coronagraphic images
due to the presence of the mask support or the diffraction spikes have
been considered in our detection performances and simulations.

\end{enumerate}

%________________________________
\subsection{Statistical results}

%________________________________
\subsubsection{Extrapolating radial velocity distributions}

\begin{figure}[t]
\centering
\includegraphics[width=\columnwidth]{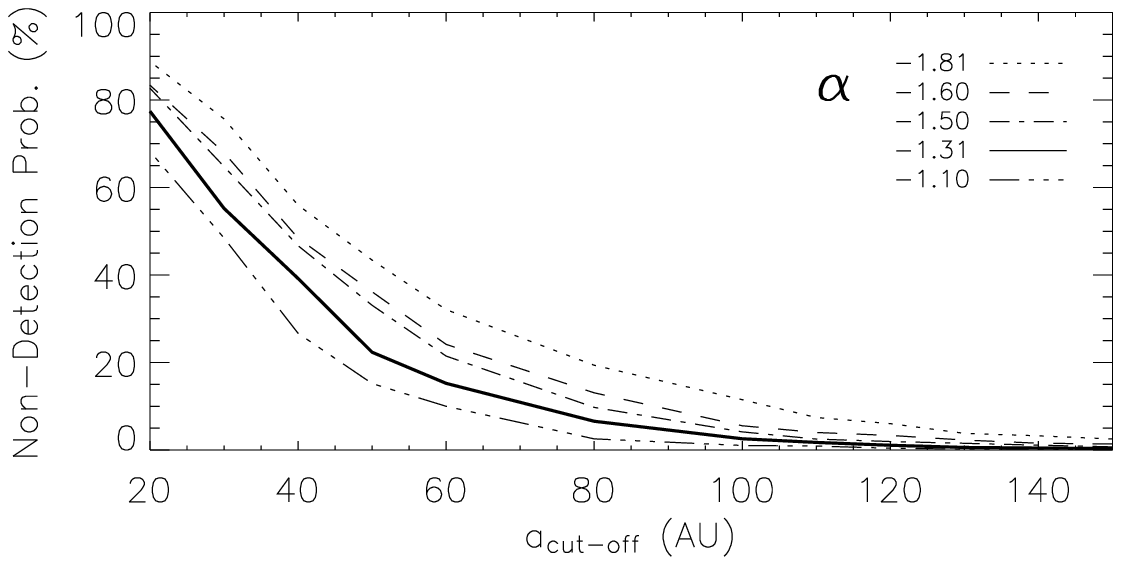}
\includegraphics[width=\columnwidth]{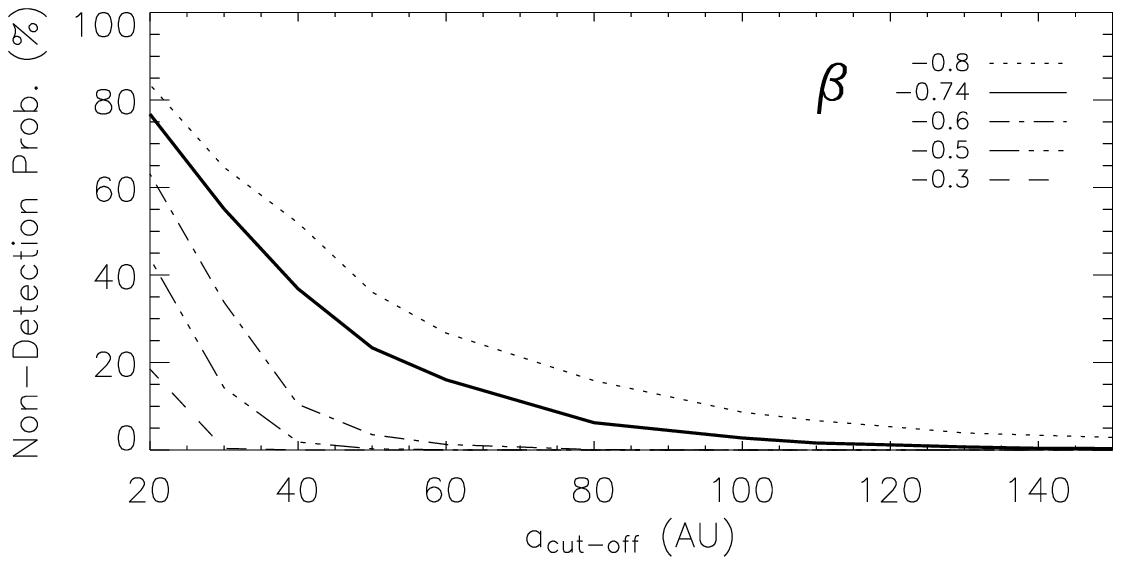}
\includegraphics[width=\columnwidth]{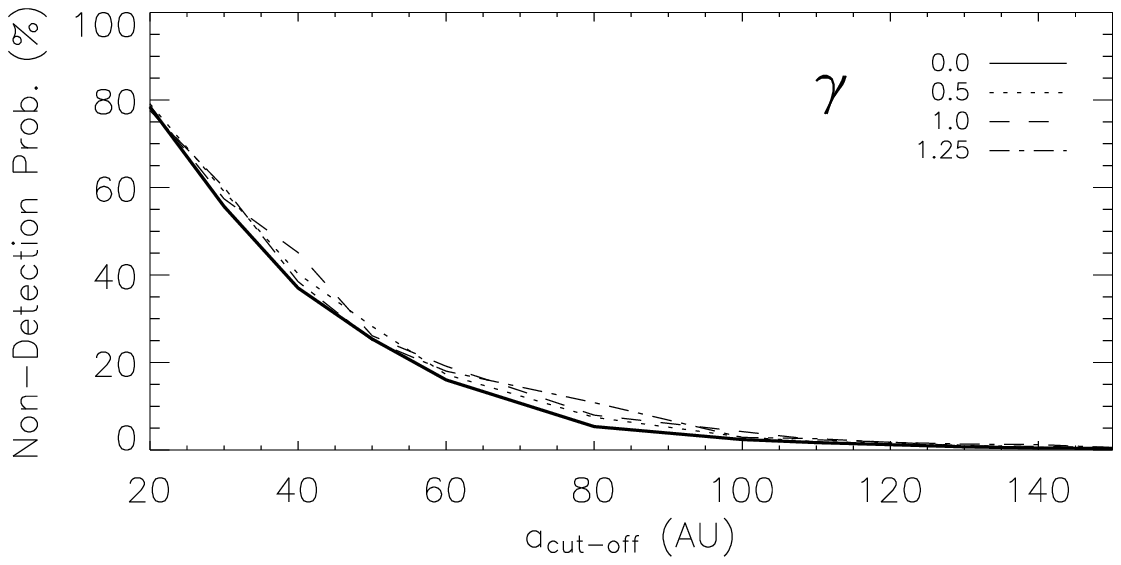}

\caption{Non-detection probability for our survey, based on various
sets of period and mass distributions as a function of the semi-major
axis cut-off of the period distribution. Mass and period distributions
are extrapolated and normalized from RV studies. \textit{Top:}
Variation of the non-detection probability with $\alpha$ and fixing
$\beta=-0.74$ and $\gamma=0.0$. \textit{Middle:} Variation of the
non-detection probability with $\beta$ and fixing $\alpha=-1.31$ and
$\gamma=0.0$. \textit{Bottom:} Variation with $\gamma$ a planet mass
scaling with the primary mass and fixing $\alpha=-1.31$ and
$\beta=-0.74$.}

\label{fig:ResSimuDistri}
\end{figure}

As a starting point, we used the mass and period distributions derived
by Cumming et al. (2008) with $\alpha=-1.31$ and $\beta=-0.74$. We
considered a giant planet frequency of $8.5\%$ in the range
$0.3-15$~\Mjup\ for periods less than 1986~days ($\leq3$~AU for a
1~\Msun\ host star). The resulting value is consistent with RV studies
of Marcy et al. (2005). Running several sets of simulations, we
explored independently the influence of period, planet mass and
primary mass distributions on the non-detection probability determined
as a function of the period cut-off. The period cut-off was chosen to
correpond to a semi-major axis cut-off between 20 and 150~AU.  The
results are reported in Fig.~\ref{fig:ResSimuDistri}, to study the
impact of the planet mass power law index $\alpha$ with $\beta=-0.74$
and $\gamma=0.0$ (\textit{Top}), of $\beta$ the period power law index
with $\alpha=-1.31$ and $\gamma=0.0$ (\textit{Middle}), and the
evolution implied by a planet mass dependency with the primary mass
when $\gamma$ varies and $\alpha=-1.31$ and $\beta=-0.74$
(\textit{Bottom}). As reference, Cumming et al. (2008) extrapolated
distributions are reported in \textit{thick solid} lines in all panels
of Fig.~\ref{fig:ResSimuDistri}. As a result, the non-detection
probability of our survey as a function of the period cut-off is more
sensitive to the variation of $\beta$, the period power law index. A
reasonable set of values can significantly be excluded for large
semi-major axis cut-off. In comparison, the influence of $\alpha$ and
$\gamma$ remains limited under the current assumptions.

%________________________________
\subsubsection{Exoplanet fraction upper limit}

The probability of planet detection for a survey of $N$ stars is
described by a binomial distribution, given a success probability
$fp_j$ with $f$ the fraction of stars with planets and $p_j$ the
individual detection probabilities of detecting a planet if present
around the star j. In our case, we can consider a null detection
result and replace each individual $p_j$ by $\langle p_j \rangle$ the
mean survey detection probability of detecting a planet if
present. Finally, assuming that the number of expected detected
planets is small compared to the number of stars observed ($f \langle
p_j \rangle << 1$), the binomial distribution can be approximated by a
Poisson distribution to derive a simple analytical solution for the
exoplanet fraction upper limit $f_{\rm{max}}$ for a given level of
credibility $\rm{CL}$.

\begin{equation}
\hspace{2.6cm}f_{\rm{max}} = \frac{-\rm{ln}(1-\rm{CL})}{N \langle p_j \rangle}
\end{equation}

\begin{figure}[t]
\centering
\includegraphics[width=\columnwidth]{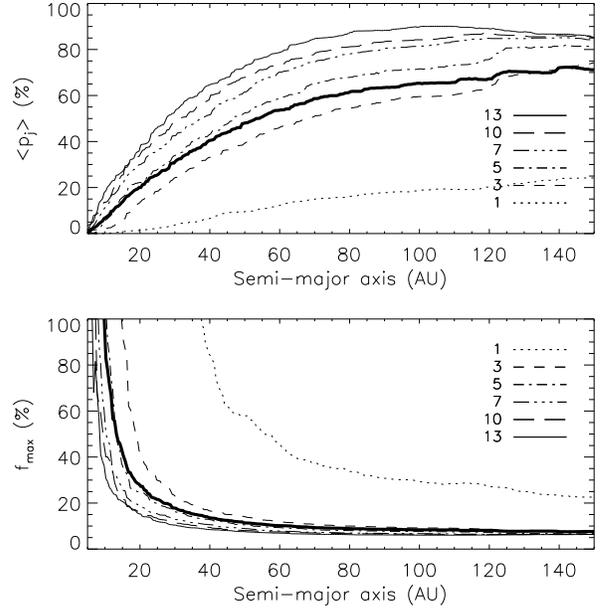}

\caption{\textit{Top:} Survey mean detection probability derived as a
function of semi-major axis assuming parametric mass and period
distributions derived by Cumming et al. (2008), i.e with
$\alpha=-1.31$, $\beta=-0.74$ and $\gamma=1.25$. The results are
reported for individual masses: 1, 3, 5, 7, 10 and 13~\Mjup. The
integrated probability for the planetary mass regime is shown with the
\textit{thick solid} line. \textit{Bottom:} Planet fraction upper
limit derived as a function of semi-major axis, given the same
mass and period distributions.}

\label{fig:ResSimuPlantFrac}
\end{figure}

We consider the period and mass power law indexes from Cumming et
al. (2008) $\alpha=-1.31$, $\beta=-0.74$ and $\gamma=1.25$ for the
period and mass distribution of giant planet. For the set of detection
limits of our survey, we can then determine $\langle p_j \rangle$, the
survey mean probability of detecting a planet if present around each
star of our sample. Then, given a confidence level $\rm{CL}=0.95$, we
obtain $f_{\rm{max}}$ as a function of planet mass and semi-major
axis. The survey mean detection probability and $f_{\rm{max}}$ are
reported in Fig.~\ref{fig:ResSimuPlantFrac}.  It is important to note
that both results depend on the assumed (mass, period, eccentricity)
distributions of the giant planet population. Similar to other deep
imaging surveys, our study begins to constrain the fraction of stars
with giant planets to less than $10\%$ for semi-major axes larger than
typically 40~AU for this specific set of period, mass and eccentricity
distributions. We also see that we barely constrain the fraction of
1~\Mjup\ planets potentially detectable for 24\% of our targets (67\%
for the 3~\Mjup\ planets). Increasing the sample size will enable
refinement of the statistical constraints on the upper limits of the
fraction of stars with giant planets as a function of their mass and
semi-major axis. However, a number of intrinsic limitations (detection
threshold, age determination and model calibration) remain that will
have to be overcome to draw more robust conclusions. Future work
gathering detection performances from multiple surveys should help
refine our knowledge of the occurence of giant planets at wide orbits
($>10$~AU) and thus complement RV survey results.

%________________________________
\subsection{Limitations}

Added to the detection threshold determination (detailed
previously), the age determination of the young, nearby stars and the
use of uncalibrated evolutionary models are the three main limitations
that directly impact the estimation of the explored planetary masses
from observed luminosities. Added to the current assumption made on
the extrapolation of close-in exoplanet distributions at wide orbits,
they limit the current relevancy of all statistical analysis of deep
imaging surveys aimed at constraining the population of giant
  planets.

%\subsubsection{Detection threshold}

%
% . actual technique if contraste estimation
%    - sliding box
%    - ring
% . Assume a guassian distributted noise and a 5sigma threshold for companion detection
% . Recent simulations show that approximation in not correct for High AO imaging
%   as quasi static speckles noise do not follow a guassain distribution. 
%   region limited by speckle noise 1.0-2.0'' (<50-100 AU), inner region 
%

\subsubsection{Age determination}

Ages of young stars near the Sun are deduced based on photometric,
spectroscopic and kinematics studies; various diagnostics are commonly
used, depending on the spectral type and age of a given star.  Details
may be found in ZS04 and T08.  In general, the most reliable ages are
obtained for stars that can be placed reliably into a moving group or
association.

Our sample is composed of 88 stars, including 51 members of known
young, nearby associations (TWA, $\beta$ Pic, Tuc-Hor and AB
Dor). Ages for the TWA and $\beta$ Pic associations have been
reasonably well constrained by various and independent (stellar
properties characterization and dynamical trace-back) studies to:
8$_{-3}^{+4}$ Myr (TWA; de la Reza et al. 2006; Barrado y Navasu\'es
2006; Scholz et al. 2007) and 12$_{-4}^{+8}$~Myr ($\beta$ Pic,
Zuckerman et al. 2001b, Ortega et al. 2004) respectively. Isochrones,
lithium depletion and X-ray luminosoties indicate an age for Tuc-Hor
of 30~Myr (Zuckerman et al. 2001a).  The age of the AB Dor association
is in some dispute (see Zuckerman et al. 2004, Luhman et al. 2005,
Luhman \& Potter 2006, Lopez-Santiago et al. 2006, Janson et al. 2007,
Ortega et al. 2007, Close et al. 2007, Boccaletti et al. 2008, Torres
et al. 2008).  In our simulations, we have assumed an age of 70 Myr
for AB Dor stars.

In our statistical analysis of 65 stars observed in coronagraphic
imaging mode, 45 are confirmed members of known associations while 17
are young candidates, currently not identified as members of any
kinematic group which makes an age estimate particularly difficult.
An excellent example of a young star not known to be a member of the
above listed moving groups is HR 8799, identified by Marois et al
(2008b) as orbited by 3 giant planets but with an age uncertainly
between 30 and 160 Myr.  In our analysis, age is directly used to
convert the detection limits to mass using evolutionary
models. Therefore, age determination remains a main limitation in this
work and others to constrain reliably the properties of a putative
population of giant planets around young, nearby stars.
 
\subsubsection{Evolutionary models}

%for studies as fundamental as the (sub)stellar
%inti mass function in young associations and the characterization of
%brown dwarf and giant planet companions imaged around nearby stars. 

Evolutionary model predictions are commonly used to infer substellar
masses from observed luminosities, as we did to convert our survey
detection performances into planetary mass limits. For stars and brown
dwarfs formed by gravitational collapse and fragmentation, models
consider the idealized description of non-accreting systems
contracting at large initial radii.  Remaining circumstellar material,
accretion and uncertainties related to choice of initial conditions
imply that comparison between observations and models are quite
uncertain at young $\le100$~Myr ages (Baraffe et al. 2002). This could
be even worse for young giant planets; the implementation of the
core-accretion mechanism as initial conditions for evolutionary
calculation could substantially change the model predictions (Marley
et al. 2007). Then massive giant planets could be significantly
fainter than equal-mass objects formed in isolation via gravitational
collapse. However, a critical issue is treatment of the accretion
shock through which most of the giant planet mass is processed and
which remains highly uncertain. In previous analyses of survey
detection performances, only predictions from Chabrier et al. (2000)
and Baraffe et al. (2003) models were used.  Use of Burrows et
al. (2003), assuming the same initial conditions, does not change
significantly the results (Nielsen et al. 2008).

%
%______________________________________________________________
\section{Conclusions}

We have conducted a deep adaptive optics imaging survey with NACO at
the VLT of 88 nearby stars of the southern hemisphere.  Our selection
criteria favored youth ($\le100$~Myr) and proximity to Earth ($\le100$~pc) to
optimize the detection of close planetary mass companions. Known
visual binaries were excluded to avoid degrading the NACO AO and/or
coronagraphic detection performances. Among our sample, 51 stars are
members of young, nearby comoving groups. 32 are young, nearby stars
currently not identified as members of any currently known association
and 5 have been reclassified as older ($\ge100$~Myr) systems. The
spectral types cover the sequence from B to M spectral types with
$19\%$ BAF stars, $48\%$ GK stars and $33\%$ M dwarfs.  The separation
investigated typically ranges between $0.1~\!''$ to $10~\!''$,
i.e. between typically 10 to 500~AU.  A sample of 65 stars was
observed in deep coronagraphic imaging to enhance our contrast
performances to $10^{-6}$ and to be sensitive to planetary mass
companions down to 1~\Mjup\ (at 24\% of our sample) and 3~\Mjup\ (at
67\%).  We used a standard observing sequence to precisely measure the
position and the flux of all detected sources relative to their visual
primary star. Repeated observations at several epochs enabled us to
discriminate comoving companions from background objects. The main
results are that:
\begin{itemize}
\item[-] we discovered of 17 new close ($0.1-5.0~\,''$) multiple
   systems. HIP\,108195\,AB and C (F1III-M6), HIP\,84642\,AB
   ($a\sim14$~AU, K0-M5) and TWA22\,AB ($a\sim1.8$~AU; M6-M6) are
   confirmed as comoving systems. TWA22\,AB, with 80\% of its orbit
   already resolved, is likely to be a rare astrometric calibrator for
   testing evolutionary model predictions.
\item[-] about $236$ faint CCs were detected around 36 stars observed
   in coronagraphy. Follow-up observations with VLT or HST for 30
   stars enabled us to identify their status. 1\% of the CCs detected
   have been confirmed as comoving companions, 43\% have been
   identified as probable background contaminants and about 56\% need
   further follow-up observations. The remaining CCs come mostly from
   the presence of crowded fields in the background of the 6 stars
   observed at one epoch.
\item[-] we confirmed previously discovered substellar companions
   around GSC\,08047-00232, AB\,Pic and 2M1207 and placed them in the
   perspective of confirmed substellar companions among young,
   nearby associations.
\item[-] finally, the statistical analysis of our complete set of
   detection limits enables us to constrain at large semi-major axes,
   20 to a few 100 AU, various mass, period and eccentricity
   distributions of giant planets extrapolated and normalized from RV
   surveys. It enables us to derive limits on the occurence of giant
   planets for a given set of physical and orbital distributions.  The
   survey starts constraining significantly the population of giant
   planet for masses $\ge3$~\Mjup. 
\end{itemize}

In the first few years following the discovery of the companion to
2M1207 (Chauvin et al. 2004), all planetary mass companions were
discovered at relatively wide separations or with small mass ratio
with their primaries. However, the recent discoveries of planetary
mass objects around the star Fomalhaut (Kalas et al. 2008), HR\,8799
(Marois et al. 2008b) and $\beta$ Pictoris (Lagrange et al. 2009), now
open a new era for the deep imaging study of giant planets that probably
formed like those of our solar system. In the perspective of
on-going and future deep imaging instruments either from the ground
(Gemini/NICI, Subaru/HiCIAO, SPHERE, GPI, EPICS) or from space (JWST,
TPF/Darwin), this work represents a pioneer successful study,
providing, with other surveys, precise information (stellar and substellar 
multiplicity,
non-detections and background contaminants) to better characterize the
overall environment of young, nearby stars, that will be prime
targets for futur exoplanets search.

% . planetray mass companions
% . recent discovery
% . future surveys..

%
%______________________________________________________________
\begin{acknowledgements}

We thank the ESO Paranal staff for performing the service mode
observations. We also acknowledge partial financial support from the
PNPS and Agence National de la Recherche, in France, from INAF through
PRIN 2006 ``From disk to planetray systems: understanding the origin
and demographics of solar and extrasolar planetary systems'' and from
NASA in the USA. We also would like to thanks France Allard and
Isabelle Baraffe for their inputs on evolutionary models and synthetic
spectral libraries. Finally, our anonymous referee for her/his
detailed and very constructive report. 

\end{acknowledgements}

%
%______________________________________________________________

\cleardoublepage

%
%__________________________________________________________________
% Tables des Contaminants/Compagnons identifies

\begin{table*}[h]

\caption{Characterization and identification (for multi-epochs
observations) of all faint sources detected during the VLT/NACO
survey. Target name, observing date and set-up are given, as well as
the different sources identified with their relative position,
relative flux and their status identification based on follow-up
observations. Sources refered as undefined (U) are objects detected at
only one epoch, (B) objects identified as stationary background
contaminants and (C) confirmed comoving companion. When VLT
observations are combined to other instrumentation (HST, USNO, 2MASS),
a flag or a reference is reported in the last column.}

\label{tab:faint_candidates}      
\centering          
\begin{tabular*}{\textwidth}{@{\excs}lllllllll}     % 7 columns 
\hline\hline       
                      % To combine 4 columns into a single one 
Name             & Candidate& UT Date        &  Filter        & Separation        & P.A.              & $\Delta m$          & Status &  Note    \\ 
                 &          &                &                & (arcsec)          & ($^{o}$)          & (mag)               &        &          \\
\hline 
GSC08047-00232   & cc-1     & 23/11/2002     & K$_s$          & $3.274\pm0.012$   & $358.85\pm0.23$   & $6.3\pm0.1$         &     &  \\
                 &          & 07/09/2003     & K$_s$          & $3.266\pm0.011$   & $358.89\pm0.23$   & $$                  &     & \\
                 &          & 05/03/2004     & K$_s$          & $3.260\pm0.012$   & $358.82\pm0.22$   & $$                  & C   &  \\
\hline 
HIP30034 (AB Pic)& cc-1     & 17/03/2003     & H              & $5.460\pm0.014$   & $175.33\pm0.18$   & $7.6\pm0.1$         &     & \\
                 &          & 05/03/2004     & H              & $5.450\pm0.016$   & $175.13\pm0.21$   & $$                  &     & \\
                 &          & 26/09/2004     & H              & $5.453\pm0.014$   & $175.10\pm0.20$   & $$                  &     & \\
\hline 
\end{tabular*}
\end{table*}

\begin{table*}[h]
\caption{Table 7 - cont}
\label{tab:faint_candidates1}      
\centering          
\begin{tabular*}{\textwidth}{@{\excs}lllllllll}     % 7 columns 
\hline\hline       
                      % To combine 4 columns into a single one 
Name             & Candidate& UT Date        &  Filter       & Separation        & P.A.              & $\Delta m$         & Status &  Note    \\ 
                 &          &                &               & (arcsec)          & ($^{o}$)          & (mag)              &        &          \\
\hline 
                 & cc-16    & 08/05/2007     & K$_{s}$       & $5.083\pm0.020$   & $81.32\pm0.22$     &  $11.3\pm0.2  $     & U & \\ 
                 & cc-17    & 08/05/2007     & K$_{s}$       & $7.208\pm0.027$   & $53.12\pm0.27$     &  $10.5\pm0.2  $     & U & \\ 
                 & cc-1     & 08/05/2007     & K$_{s}$       & $8.158\pm0.030$   & $51.12\pm0.27$     &  $11.1\pm0.2  $     & U & \\ 
                 & cc-19    & 08/05/2007     & K$_{s}$       & $7.650\pm0.029$   & $61.15\pm0.24$     &  $10.5\pm0.2  $     & U & \\ 
                 & cc-20    & 08/05/2007     & K$_{s}$       & $6.792\pm0.026$   & $80.03\pm0.21$     &  $11.1\pm0.2  $     & U & \\ 
                 & cc-21    & 08/05/2007     & K$_{s}$       & $7.705\pm0.030$   & $75.99\pm0.21$     &  $9.2\pm0.2  $      & U & \\
\hline 
\end{tabular*}
\end{table*}

\begin{table*}[h]
\caption{Contaminants and companion identified}
\label{tab:faint_candidates2}      
\centering          
\begin{tabular*}{\textwidth}{@{\excs}lllllllll}     % 7 columns 
\hline\hline       
                      % To combine 4 columns into a single one 
Name             & Candidate& UT Date        &  Filter       & Separation        & P.A.               & $\Delta m$         & Status &  Note    \\ 
                 &          &                &               & (arcsec)          & ($^{o}$)           & (mag)              &        &          \\
\hline                  
                 & cc-16    & 27/04/2004     &  H           & $6.266\pm0.019$    & $193.90\pm0.24$    &       &    & \\ 
                 &          & 26/02/2006     &  K$_s$       & $6.133\pm0.018$    & $191.73\pm0.19$    &  $10.4\pm0.2  $     & B   & \\ 
                 & cc-17    & 27/04/2004     &  H           & $3.466\pm0.017$    & $227.25\pm0.37$    &  $10.8\pm0.2  $     & PB   & HST \\ 
                 & cc-18    & 27/04/2004     &  H           & $13.912\pm0.037$   & $205.66\pm0.21$    &       &    & \\ 
                 &          & 26/02/2006     &  K$_s$       & $13.701\pm0.031$   & $204.91\pm0.16$    &  $10.2\pm0.2  $     & B   & \\ 
                 & cc-19    & 27/04/2004     &  H           & $14.827\pm0.040$   & $207.38\pm0.20$    &       &    & \\
\hline
\end{tabular*}
\end{table*}

\begin{table*}[h]
\caption{Contaminants and companion identified}
\label{tab:faint_candidates3}      
\centering          
\begin{tabular*}{\textwidth}{@{\excs}lllllllll}     % 7 columns 
\hline\hline       
                      % To combine 4 columns into a single one 
Name             & Candidate      & UT Date        &  Filter       & Separation        & P.A.              & $\Delta m$         & Status &  Note    \\ 
                 &                &                &               & (arcsec)          & ($^{o}$)          & (mag)              &        &          \\
\hline 
                 & cc-4     & 17/03/2003     &  H           & $4.691\pm0.017$   & $66.56\pm0.18$    &      &    & \\ 
                 &          & 27/04/2004     &  H           & $4.763\pm0.019$   & $66.30\pm0.24$    &  $10.3\pm0.2  $     & B   & \\
                 & cc-5     & 17/03/2003     &  H           & $6.891\pm0.025$   & $70.95\pm0.15$    &      &    & \\ 
                 &          & 27/04/2004     &  H           & $6.971\pm0.027$   & $70.51\pm0.22$    &  $11.2\pm0.2  $     & B   & \\ 
                 & cc-6     & 17/03/2003     &  H           & $3.166\pm0.014$   & $97.83\pm0.16$    &       &    & \\ 
                 &          & 27/04/2004     &  H           & $3.229\pm0.014$   & $96.97\pm0.23$    &  $9.3\pm0.2  $     & B   & \\
\hline 
\end{tabular*}
\end{table*}

\begin{table*}[h]
\caption{Contaminants and companion identified}
\label{tab:faint_candidates4}      
\centering          
\begin{tabular*}{\textwidth}{@{\excs}lllllllll}     % 7 columns 
\hline\hline       
                      % To combine 4 columns into a single one 
Name             & Candidate& UT Date        &  Filter        & Separation        & P.A.              & $\Delta m$          & Status &  Note    \\ 
                 &          &                &                & (arcsec)          & ($^{o}$)          & (mag)               &        &          \\
\hline 
HIP88399         & cc-1     & 24/09/2004     &  K$_{s}$       & $11.242\pm0.028$  & $17.93\pm0.21$     &  $13.3\pm0.2  $    & PB   & HST \\ 
                 & cc-2     & 24/09/2004     &  K$_{s}$       & $13.045\pm0.038$  & $36.55\pm0.22$     &  $15.\pm0.2  $     & U   & \\ 
                 & cc-3     & 24/09/2004     &  K$_{s}$       & $11.155\pm0.034$  & $46.37\pm0.24$     &  $14.6\pm0.2  $    & U   & \\ 
                 & cc-4     & 24/09/2004     &  K$_{s}$       & $14.598\pm0.043$  & $138.5\pm0.23$     &  $12.\pm0.2  $     & U   & \\ 
                 & cc-5     & 24/09/2004     &  K$_{s}$       & $14.774\pm0.043$  & $138.87\pm0.22$    &  $11.6\pm0.2  $    & U   & \\ 
                 & cc-6     & 24/09/2004     &  K$_{s}$       & $9.735\pm0.023$   & $178.36\pm0.22$    &  $13.9\pm0.2  $    & U   & \\
\hline 
\end{tabular*}
\end{table*}

\begin{table*}[h]
\caption{Contaminants and companion identified}
\label{tab:faint_candidates5}      
\centering          
\begin{tabular*}{\textwidth}{@{\excs}lllllllll}     % 7 columns 
\hline\hline       
                      % To combine 4 columns into a single one 
Name             & Candidate      & UT Date        &  Filter       & Separation        & P.A.              & $\Delta m$         & Status &  Note    \\ 
                 &                &                &               & (arcsec)          & ($^{o}$)          & (mag)              &        &          \\
\hline 
                 & cc-13     & 19/08/2005     &  K$_{s}$       & $10.2140\pm0.024$   & $20.55\pm0.20$    &  $12.6\pm0.2  $     & U   & \\ 
                 & cc-14     & 19/08/2005     &  K$_{s}$       & $15.5660\pm0.109$   & $40.75\pm0.47$    &  $13.6\pm0.2  $     & U  & \\ 
                 & cc-15     & 19/08/2005     &  K$_{s}$       & $15.7480\pm0.043$   & $52.83\pm0.21$    &  $12.6\pm0.2  $     & U  & \\ 
                 & cc-16     & 19/08/2005     &  K$_{s}$       & $10.6060\pm0.023$   & $74.40\pm0.20$    &  $12.6\pm0.2  $     & U  & \\ 
                 & cc-17     & 19/08/2005     &  K$_{s}$       & $7.913\pm0.015$   & $84.50\pm0.20$    &  $12.4\pm0.2  $     & U  & \\ 
                 & cc-18     & 19/08/2005     &  K$_{s}$       & $8.452\pm0.023$   & $117.150\pm0.2$    &  $13.3\pm0.2  $     & U  & \\
\hline 
\end{tabular*}
\end{table*}

\begin{table*}[h]
\caption{Contaminants and companion identified}
\label{tab:faint_candidates6}      
\centering          
\begin{tabular*}{\textwidth}{@{\excs}lllllllll}     % 7 columns 
\hline\hline       
                      % To combine 4 columns into a single one 
Name             & Candidate      & UT Date        &  Filter       & Separation        & P.A.              & $\Delta m$         & Status &  Note    \\ 
                 &                &                &               & (arcsec)          & ($^{o}$)          & (mag)              &        &          \\
\hline
FS1017           & cc-1     & 19/08/2005     &  K$_{s}$       & $12.041\pm0.032$   & $57.93\pm0.20$    &  $4.8\pm0.1  $     & U   & \\ 
                 & cc-2     & 19/08/2005     &  K$_{s}$       & $9.463\pm0.025$   & $213.830\pm0.20$    &  $5.9\pm0.1  $     & U   & \\ 
                 & cc-3     & 19/08/2005     &  K$_{s}$       & $9.689\pm0.024$   & $243.940\pm0.20$    &  $4.1\pm0.1  $     & U   & \\ 
                 & cc-4     & 19/08/2005     &  K$_{s}$       & $13.799\pm0.038$   & $318.650\pm0.21$    &  $6.3\pm0.1  $     & U   & \\ 
                 & cc-5     & 19/08/2005     &  K$_{s}$       & $15.149\pm0.042$   & $317.320\pm0.22$    &  $6.1\pm0.1  $     & U   & \\ 
                 & cc-6     & 19/08/2005     &  K$_{s}$       & $14.507\pm0.040$   & $119.520\pm0.21$    &  $7.5\pm0.2  $     & U   & \\
\hline 
\end{tabular*}
\end{table*}

\end{document}